\documentclass[useAMS,usenatbib]{mn2e}

\usepackage{verbatim}
\usepackage{graphicx}
\usepackage{amssymb}
\usepackage{amsmath}
\usepackage{xspace}

\newcommand{\ro}[1]{\ensuremath{\textrm{#1}}}

\newcommand{\kmsec}{\ensuremath{~\ro{km}~ \ro{s}^{-1}}}
\newcommand{\ergs}{\ensuremath{~\ro{erg s}^{-1}}}
\newcommand{\cm}{\ensuremath{~\ro{cm}^{-2}}}
\newcommand{\Msol}{\ensuremath{M_{\odot}}\xspace}
\newcommand{\lya}{\ensuremath{\ro{Ly}\alpha}\xspace}
\newcommand{\ps}{Press-Schechter }

\newcommand{\hi}{\ensuremath{\ro{H\textsc{i}}}\xspace}
\newcommand{\nhi}{\ensuremath{N_{\ro{H\textsc{i}}}}\xspace}
\newcommand{\dhi}{\ensuremath{n_{\ro{H\textsc{i}}}}\xspace}
\newcommand{\nuv}{\ensuremath{\hat{\boldsymbol{n}}}\xspace}
\newcommand{\df}{\ensuremath{~ \ro{d} }}
\newcommand{\dd}{\ensuremath{\ro{d} }}
\newcommand{\rv}{\ensuremath{\boldsymbol{r}}\xspace}
\newcommand{\vv}{\ensuremath{\boldsymbol{v}}\xspace}
\newcommand{\DnuL}{\ensuremath{\Delta \nu_{\ro{L}}}\xspace}
\newcommand{\DnuD}{\ensuremath{\Delta \nu_{\ro{D}}}\xspace}
\newcommand{\xcr}{\ensuremath{x_\ro{crit}}\xspace}

\newcommand{\cN}{\ensuremath{\mathcal{N}}\xspace}

\defcitealias{2006ApJ...649...14D}{DHS06}
\defcitealias{2008ApJ...681..856R}{R08}
\defcitealias{2009MNRAS.397..511B}{BH09}
 
\title[Extended \lya emission from DLAs]{Faint extended \lya emission due to star formation 
 at the centre of high column density QSO absorption systems}

\author[Barnes \& Haehnelt]{Luke A. Barnes\thanks{E-mail:
lab@ast.cam.ac.uk (LAB); haehnelt@ast.cam.ac.uk (MGH)} and Martin G. Haehnelt\\ 
Institute of Astronomy and Kavli Institute for Cosmology, Madingley Road, Cambridge, CB3
0HA}

\begin{document}

\date{not yet submitted}

\pagerange{\pageref{firstpage}--\pageref{lastpage}} \pubyear{2008}

\maketitle 

\label{firstpage}

\begin{abstract} 
We use detailed \lya radiative transfer calculations to further 
test the claim of Rauch et al. (2008) that they have detected spatially
extended faint \lya emission from the elusive host population of 
Damped \lya Absorption systems (DLAs) in their recent ultra-deep 
spectroscopic survey. We investigate the spatial and
spectral distribution of \lya emission due to star-formation at the
centre of DLAs, and its dependence on the spatial and velocity structure 
of the gas. Our model simultaneously reproduces the observed
properties of DLAs and the faint \lya emitters, including the 
velocity width and column density distribution of DLAs and 
the large spatial extent of the emission of the faint emitters. 
Our modelling confirms previous suggestions that DLAs are predominately
hosted by Dark Matter (DM) halos in the mass range $10^{9.5}- 10^{12} \Msol$, 
and are thus of significantly lower mass than those inferred for $L_{*}$
Lyman Break Galaxies (LBGs). Our modelling suggests that DM halos hosting DLAs 
retain up to 20\% of the cosmic baryon fraction in the form of
neutral hydrogen, and that star formation at the centre of the halos
is responsible for the faint \lya emission. The scattering of a significant fraction
of the \lya emission to the observed radii, which can be as large as 50kpc or 
more, requires the amplitude of the bulk motions of the gas at
the centre of the halos to be moderate. The observed space density and size
distribution of the emitters together with the incidence rate of DLAs
suggests that the \lya emission due to star formation has
a duty cycle of $\sim 25\%$. 
\end{abstract}


\section{Introduction} 

\citet[][hereafter \citetalias{2008ApJ...681..856R}]{2008ApJ...681..856R} 
recently reported the results of an
ultra-deep spectroscopic survey for low surface brightness 
Ly$\alpha$ emitters at redshift $z \sim 3$. A 92 hour long exposure with
the ESO VLT FORS2 instrument yielded a sample of 27 faint line emitters
with fluxes of a few times $10^{-18} $ erg s$^{-1}$cm$^{-2}$, which
they argue are likely to be dominated by \lya. They 
further conclude that the large comoving number density, $3 \times
10^{-2} ~ h^{3}_{70} ~ {\rm Mpc}^{-3}$, and the large covering factor 
$dN/dz \sim 0.2-1$ suggest that the emitters can be identified with
the elusive host population of damped Ly$\alpha$ systems (DLAs) 
and high column density Lyman limit systems. 
\citet[][hereafter \citetalias{2009MNRAS.397..511B}]{2009MNRAS.397..511B}, 
building on the successful model for DLAs of 
\citet{1998ApJ...495..647H,2000ApJ...534..594H}, presented a 
simple model that simultaneously accounts for
the kinematic properties and incidence rate of the observed
DLAs \emph{and} the luminosity function and the size
distribution of the \citetalias{2008ApJ...681..856R} emitters in the context of the
$\Lambda$CDM model for structure formation. The model assumes a simple relation 
between the size of the damped absorption and \lya emission regions, and proposes that 
the \lya luminosity is proportional to the total halo mass. 
\citetalias{2009MNRAS.397..511B} further 
corroborated the suggestion that cooling radiation is not expected to
contribute significantly to the observed \lya emission, and that the
emitters are most likely powered by star formation. In the 
model, DLAs are small galaxies hosted by DM halos with masses in 
the range $10^{9.5}$ to $10^{12}$ \Msol and have rather large low surface brightness \lya 
halos which extend to radii of up to 50kpc or larger. 

In order to fit the observed size distribution of the faint \lya
emitters, \citetalias{2009MNRAS.397..511B} assumed that the \lya
emission extends to radii somewhat larger than is required 
to reproduce the incidence rate for DLAs. However, no modelling of the gas
distribution and \lya radiative transfer was done. 
We present such modelling here to investigate whether the sizes, 
surface brightness profiles and spectral line shapes 
can be reproduced with simple but plausible assumptions for 
the distribution and the physical properties of the gas in the DM halos 
suggested by our previous modelling to be the hosts of the faint emitters. 

Studies of the radiative transfer of \lya have a long history. The
first investigations employed approximate calculations and simple
physical arguments \citep{1949BAN....11....1Z,1952PASJ....4..100U,
1962ApJ...135..195O,1971ApJ...168..575A,1972ApJ...174..439A}. Later,
analytic solutions were found for simple geometries in the limit of
large optical depth
\citep{1973MNRAS.162...43H,1990ApJ...350..216N}. It
was realised early, however, that Monte Carlo simulations provided the
most flexible method for investigating arbitrary geometry, density
distributions and velocity structures. Many investigations have
employed such techniques - 
\citet{1968ApJ...152..493A,1968ApJ...153..783A,1972ApJ...176..439C,1973A&A....24..219P,1979ApJ...233..649B,1986A&A...158..310N,2000JKAS...33...29A,2001ApJ...554..604A,2002ApJ...567..922A,2002ApJ...578...33Z,2005ApJ...628...61C,2006MNRAS.367..979H,2006ApJ...645..792T}, \citet[][hereafter
\citetalias{2006ApJ...649...14D}]{2006ApJ...649...14D}, \citet{2006A&A...460..397V,2009ApJ...696..853L}. 

Our modelling is most similar to that of \citetalias{2006ApJ...649...14D} and 
\citet{2006A&A...460..397V}, who modelled \lya radiative transfer 
in collapsing proto-galaxies and high-redshift galaxies respectively. In many 
instances we make similar assumptions as these authors, but our code 
was developed independently.

The paper is structured as
follows. In Section 2 we give a brief summary of the salient features of
our Monte-Carlo code for \lya radiative transfer and show results
for standard test problems. In Section 3 we discuss our assumptions
for the distribution and the physical properties of the gas. We also show
the dependence of the surface brightness profile and the spectral line
shapes on these assumptions. In Section 4 we present the results for
a consistent model of the size distribution and the luminosity
function of the faint \lya emitters. The technical details of the 
implementation of \lya radiative transfer Monte-Carlo algorithm are
described in an appendix. 


\section{The Monte-Carlo radiative transfer code}

\subsection{General properties of the code}
\lya is a resonant transition in hydrogen. The optical depth at line centre
$\tau_0$ of a typical \hi region is thus generally very large. 
\lya photons, which are expected to be produced
copiously in star-forming galaxies, will therefore undergo many
scatterings. Escape from the \hi region usually requires diffusion in both frequency and space. 

The properties of the emergent radiation depend sensitively on not only the spatial
distribution of the gas, but also on its kinematics, ionization state,
temperature and dust content. With modern computers Monte-Carlo
sampling of the diffusion process has become the method of choice. 
We employ such a code for simple spatial and kinematical 
configurations.

Our code makes a number of well-tested assumptions and approximations. Throughout, we
will use a dimensionless frequency variable $x$, the frequency
displacement from line centre in units of the Doppler frequency width 
(see appendix A for details). As the photons experience a large number of scatterings
before they escape, we can inject all our photons
at line centre in the fluid frame of the gas ($x_\ro{initial} = 0$)
for simplicity. The photon scattering is partially coherent --- in the rest frame of the
scattering atom, the final frequency of the photon differs from the
initial frequency only by the (often negligible) effect of atomic
recoil. The direction of the photon post-scattering is chosen from a 
dipole distribution, although choosing an isotropic distribution 
makes little difference. We also incorporate a cosmic abundance of
deuterium (see \citetalias{2006ApJ...649...14D}). 
Finally, we give all results in the rest frame of the centre of mass of the emitter.

Monte Carlo \lya RT codes like the one used here can be significantly accelerated
by skipping scatterings in the core of the line-profile
[\citet{2002ApJ...567..922A}, \citetalias{2006ApJ...649...14D}]. In
the Doppler core, the optical depth is so large that the photon experiences very little 
spatial diffusion while it awaits an encounter with a rapidly moving
atom that will scatter it in frequency space into the wings of the
line profile. The gas distributions we consider here are spherically symmetric. 
We exploit this symmetry by using an arrangement of spherical shells. 

A detailed description of the implementation and relevant formulae
of our code can be found in Appendix \ref{A:lyaRT}.

\subsection{Testing the code}
There are a number of analytical solutions to \lya radiative 
transfer problems against which we have tested our code. 

\begin{figure*} \centering
	\begin{minipage}[c]{0.48\textwidth} \centering 
 		\includegraphics[width=\textwidth]{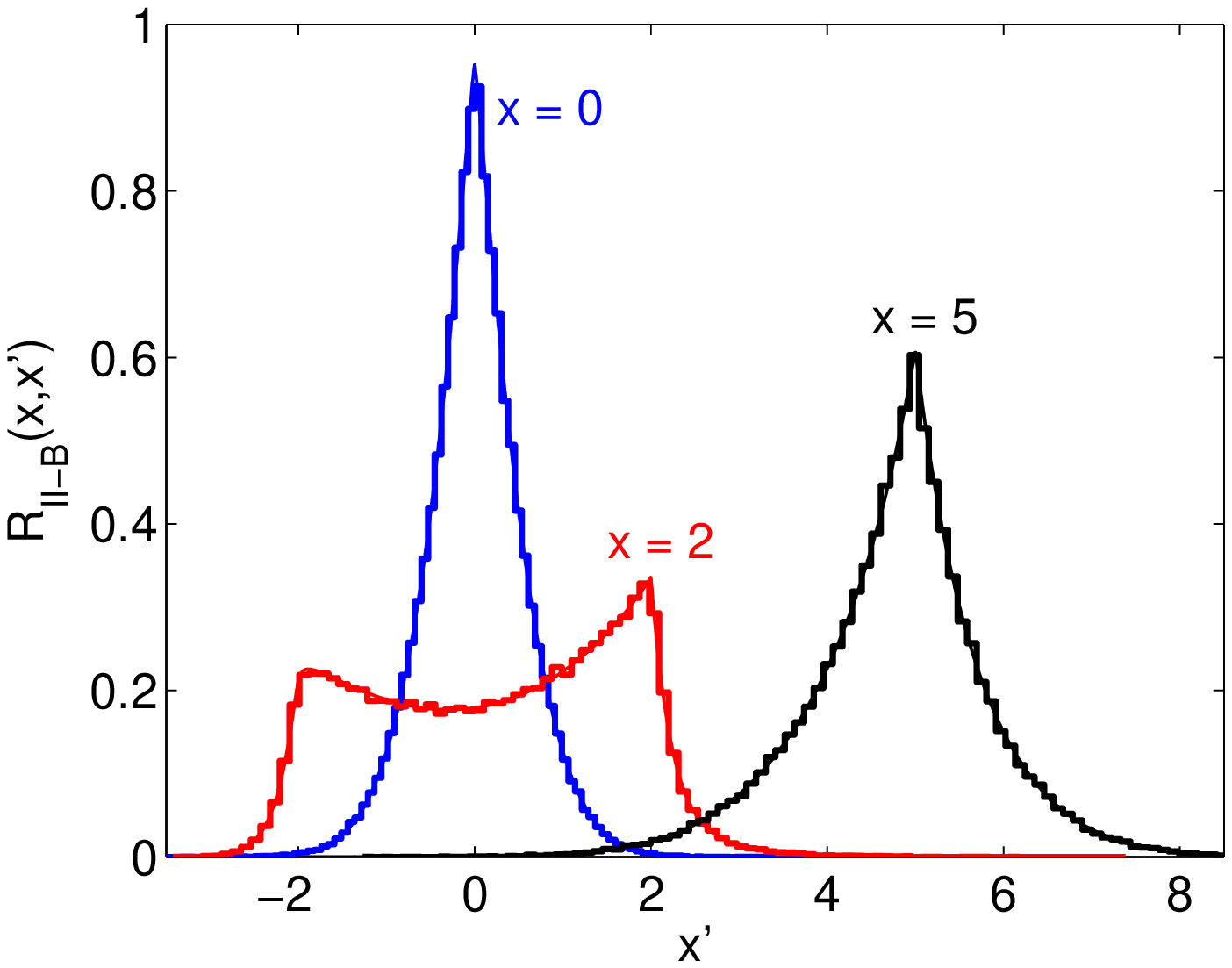}
	\end{minipage}
	\begin{minipage}[c]{0.48\textwidth} \centering 
 		\includegraphics[width=\textwidth]{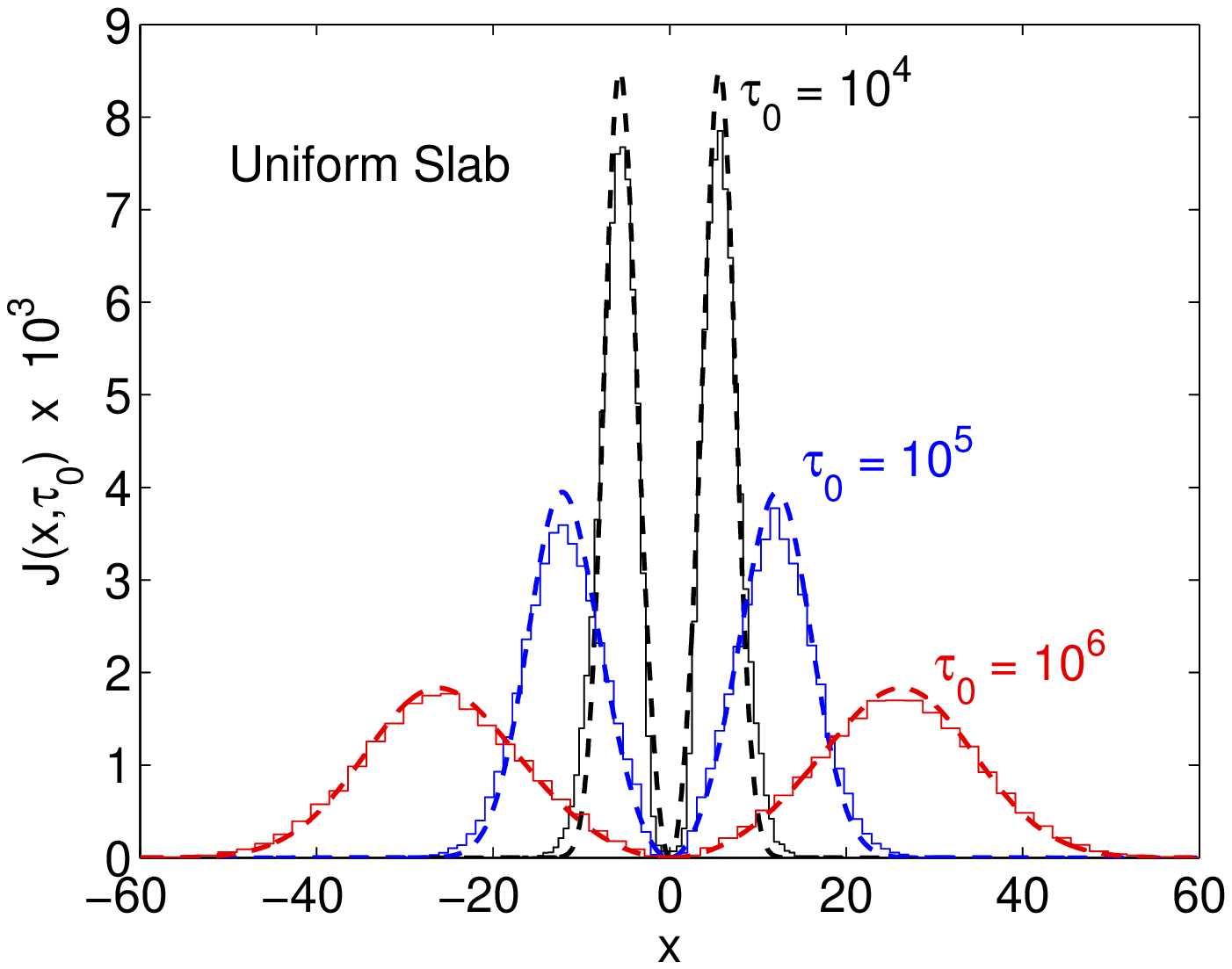}
	\end{minipage}
	\begin{minipage}[c]{0.48\textwidth} \centering 
 		\includegraphics[width=\textwidth]{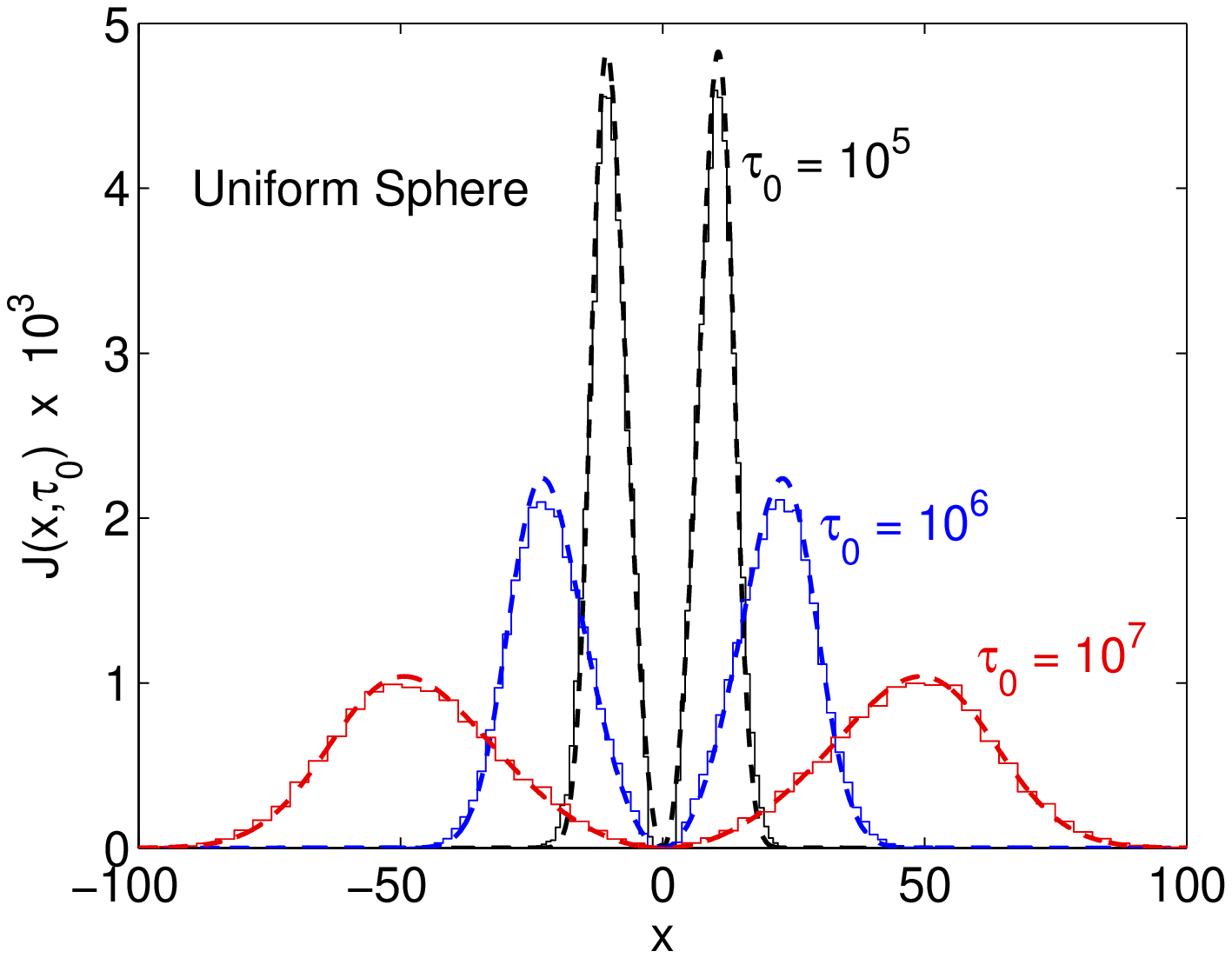}
	\end{minipage}
	\begin{minipage}[c]{0.48\textwidth} \centering 
 		\includegraphics[width=\textwidth]{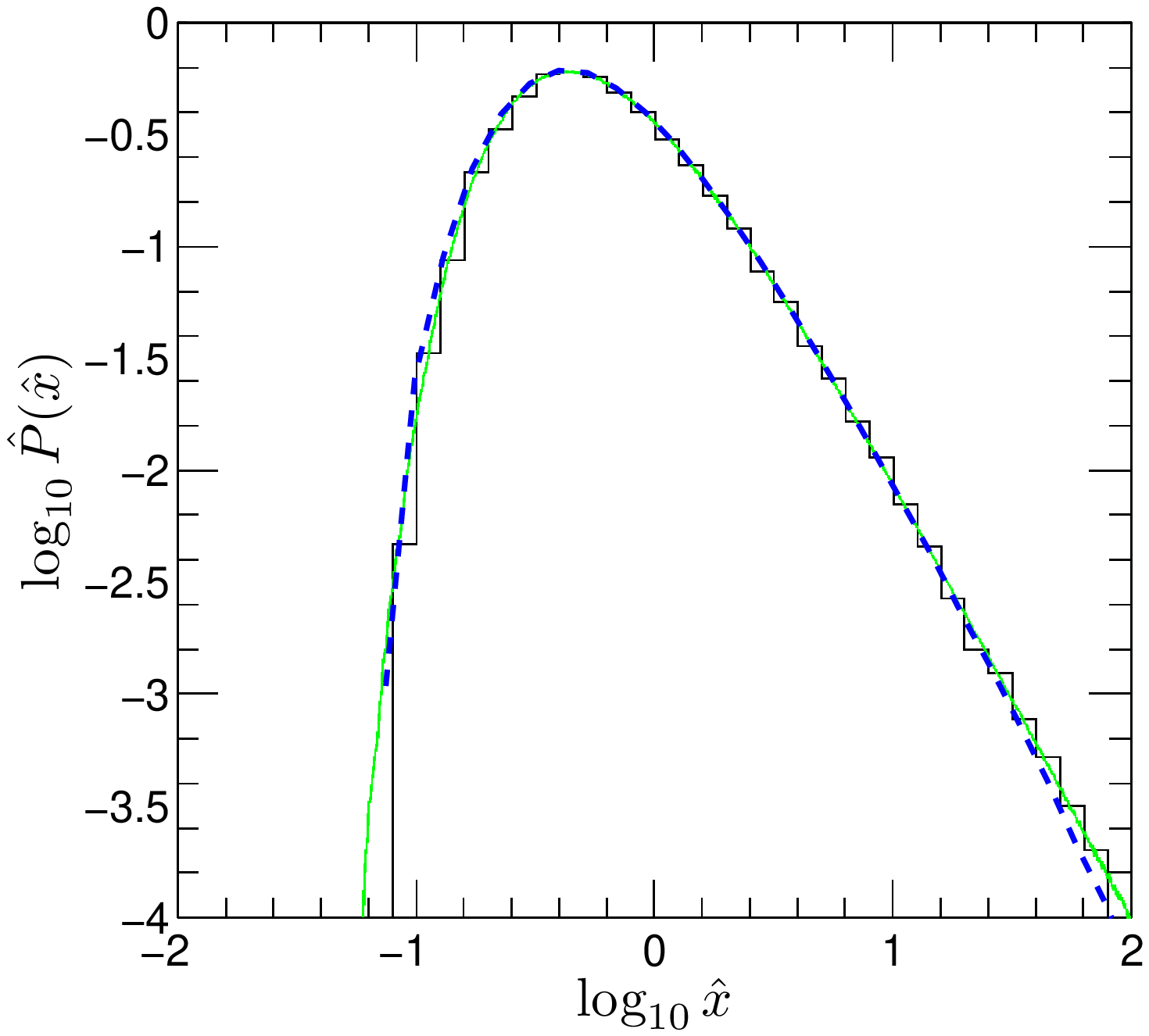}
	\end{minipage}
 \caption{Four tests of our \lya RT code. \emph{Top left panel:} The
redistribution function $R_\ro{II-B}(x,x')$ gives the probability that
a photon, whose frequency prior to absorption was $x$, is re-emitted
with frequency in the range $[x',x'+\dd x']$. The thin, smooth curve is
the analytic solution and the thick histogram is the output from our
code for $x = 0,2,5$ (blue, red, black respectively), and $T = 10$ K
($a = 0.0149$). The analytic solution and our Monte-Carlo results 
are almost indistinguishable. \emph{Top
right panel:} For an optically thick ($\sqrt{\pi}\tau_0 \gtrsim
10^3/a$), uniform, static slab of neutral hydrogen, where line-centre
photons are injected at the centre of the slab, we can compare with
the analytic solution of \citet{1973MNRAS.162...43H} and
\citet{1990ApJ...350..216N}. We set $T = 10$ K, and $\tau_0$ is the
line-centre, centre-to-edge optical depth as labeled on the plot. 
The solid histogram from our code agrees again well with the
analytical solution, especially as $\tau_0$ increases. \emph{Bottom
left panel:} As above, but for a uniform sphere. The agreement is very good. 
\emph{Bottom right panel:} \lya RT through uniform \hi
undergoing Hubble expansion with velocity $\vv_\ro{b} = H(z) \rv$. The histogram is
the result from the code of \citetalias{2006ApJ...649...14D}, the
green solid curve is the solution of by LR99 
\citet{1999ApJ...524..527L}, and the dashed blue line is the result of
our code. The agreement is again excellent. }
 \label{fig:plottest}
\end{figure*}

\citet{1962MNRAS.125...21H} calculated the redistribution function for
the case of coherent scattering in the atom's frame with radiation
damping (i.e. incorporating the Lorentzian natural line width). The
result (Equation 3.12.2 of \citet{1962MNRAS.125...21H}) is
$R_\ro{II-B}(x,x') \df x'$, which is defined as the probability that a photon,
whose frequency prior to absorption was $x$, is re-emitted with
frequency in the range $[x',x'+\dd x']$. 
In the top left panel of Figure \ref{fig:plottest}
we compare the analytic formula with the output of our code. 
The thin, smooth line is the analytic solution and the thick histogram
is the output from our code for $x = 0,2,5$ (blue, red, black
respectively) for $T = 10$ K ($a = 0.0149$). The two lines are almost 
indistinguishable.

\citet{1973MNRAS.162...43H} and \citet{1990ApJ...350..216N} derived an
analytic expression for the spectrum $J(x,\tau_0)$ of radiation emerging
from an optically thick ($\sqrt{\pi}\tau_0 \gtrsim 10^3/a$), uniform,
static slab of neutral hydrogen, where line-centre photons are
injected at the centre of the slab, atomic recoil is neglected and
$\tau_0$ is the line-centre, centre-to-edge optical depth. The
solution normalised to $(4 \pi)^{-1}$ is given by, 
\begin{equation} J(x,\tau_0) = \frac{\sqrt{6}}{24\sqrt{\pi}}
\frac{x^2}{a \tau_0} ~ \frac{1}{\cosh \left[ \sqrt{\pi^3/54} ~ |x^3|/a
\tau_0 \right]}.
\end{equation} The comparison of our code with this analytic 
solution is shown for $T = 10$K in 
the top right panel of Figure \ref{fig:plottest}. The agreement is
again very good, especially as $\tau_0$ increases.

Next we compare the results of our code to the analogue of the above solution for a uniform,
static sphere \citepalias[][Equation (9)]{2006ApJ...649...14D}. As the
bottom left panel of Figure \ref{fig:plottest} shows,
the agreement is also very good.

To test the code for the case of bulk motions in the gas, we compare our 
code with the modelling of \citet{1999ApJ...524..527L}. Loeb \& Rybicki 
investigated \lya RT for a uniform \hi distribution undergoing Hubble 
expansion with $\vv_\ro{b} = H(z) \rv$, where $H(z)$
is the Hubble constant at redshift $z$. For photons blueward of
line-centre in the fluid frame, the expansion of the universe
will eventually redshift the photon back into resonance. 
Only photons redward of line centre can propagate to
infinity. The optical depth to infinity is
\begin{equation} \tau_\infty(\bar{x}) =
-\frac{n_\hi~\sigma_0~a~v_{\ro{th}}}{\sqrt{\pi} H(z) \bar{x}} \equiv
\frac{x_{*}}{\bar{x}}, 
\end{equation} where $\bar{x}$ is the frequency in the fluid frame. 
This defines a critical frequency $x_{*}$. Photons with $\bar{x} \ll
x_{*}$ are redshifted enough to stream freely. We follow here
\citet{1999ApJ...524..527L} and use a new frequency variable
$\hat{x} = \bar{x} / x_{*}$. We refer the reader to Appendix B2 of
\citetalias{2006ApJ...649...14D} for a careful discussion of the
modifications necessary to permit a meaningful comparison between the
two codes.

In the bottom right panel of Figure \ref{fig:plottest}
we compare the results of our code with those of \citet{1999ApJ...524..527L},
and \citetalias{2006ApJ...649...14D}. The relevant parameters are:
$z = 10$, $n_\hi = 2 \times 10^{-7} (1+z)^3$ cm$^{-3}$ and $T = 10$K. Note that $\int
\hat{P}(\hat{x}) \df \hat{x} = 1$. The agreement is again excellent.


\section{\lya radiative transfer in DLAs / faint \lya emitters} 
\subsection{A simple spherically symmetric model for the spatial distribution and 
kinematics of the gas} \label{S:gasmodel}

\subsubsection{Observational and theoretical constraints}

Our knowledge of the spatial distribution and kinematics of the
neutral hydrogen in the DLAs / faint \lya emitters is still somewhat limited. 
The statistics of the occurrence DLAs and their column density
distribution give us integral constraints on 
the spatial distribution of the gas, while the velocity distribution of
low ionization species tracing neutral hydrogen gives us some
indication of the velocity range of bulk motions. The bulk motions of
the gas appear to have velocities that range from 
a few tens of km/s to several hundred km/s 
\citep{1997ApJ...487...73P}. The relative
contribution of ordered and random motions and the role of gas inflow
and galactic winds is, however, still very uncertain 
\citep{1998ApJ...495..647H,2000ApJ...534..594H,2008ApJ...683..149R,
2008MNRAS.390.1349P,2009MNRAS.397..411T}. 

If the identification of the faint \lya emitters as DLA host galaxies
is indeed correct, then this gives us for the first time constraints 
on the spatial extent of the gas distribution in individual objects
for a sizable (\lya) emission selected sample. While there is still
significant ambiguity due to the unknown ``duty cycle'' for \lya emission 
(as discussed by \citetalias{2009MNRAS.397..511B}),
theoretical modelling gives us a handle on the halo masses and virial velocities 
expected to host the DLAs / faint \lya emitters. By duty cycle, we
mean the fraction of DM host haloes detectable through their \lya emission 
at any given time. There are a number of reasons why this fraction 
may be smaller than unity. The haloes, for example, may not continuously 
form stars, but geometrical factors due to a
preferential escape of the \lya emission in certain directions due to variations in
the neutral hydrogen column density (and/or the dust column density) 
should also contribute \citep{2009ApJ...696..853L,2009arXiv0907.2698L}.  

Given these uncertainties, we have decided to follow \citetalias{2006ApJ...649...14D}
and investigate \lya radiative transfer for sources at the 
centre of DM halos with a range of masses, a 
duty cycle for \lya emission, a simple spherically 
symmetric gas distributions without dust, and either inflow or outflow with 
velocities which vary as a power law with radius.
This already leads to a rich variety in the predicted surface
brightness profiles and spectral shapes and allows us to study the
influence of important physical parameters. 

\subsubsection{The assumed radial distribution of neutral hydrogen}
We begin by specifying the radial distribution of neutral hydrogen
in a given halo. The total amount of hydrogen is set relative to the 
cosmic\footnote{The relevant cosmological parameters used in this work are: 
$(h, \Omega_M, \Omega_b,\Omega_{\Lambda}, \sigma_8, n) = (0.7, 0.3, 0.045, 0.7, 0.9,
1)$.} mass fraction of hydrogen $f_\ro{H} = \Omega_\ro{H} / \Omega_\ro{m}$. 
Throughout we assume a helium fraction of $Y_{\rm p}=0.24$. There are a 
number of reasons to suspect that the
hydrogen mass fraction in a typical halo is lower than $f_\ro{H}$.
Firstly, baryons are subject to the smoothing effects of gas
pressure. Secondly, gas that forms stars is both ionised and extremely
compact. Stars will also ionise the neutral gas around them --- this
is a source of \lya but also reduces the amount of \hi that remains to
scatter photons. Finally, stellar and AGN driven galactic winds are
expected to drive gas out of galaxies back into the IGM.
As a first attempt at modelling this effect, we reduce the total mass
of baryons in the halo to a fraction $f_{\rm e}$ of the cosmic value.

On top of reducing the amount of neutral hydrogen in a typical halo, it is 
known that the UV background will significantly ionise the gas in 
halos too small to self-shield. It is also easier for galactic winds to 
drive gas out of small, shallow halos. In \citetalias{2009MNRAS.397..511B}, 
this effect was implemented via an exponential suppression of the 
cross-section of neutral hydrogen below a critical circular velocity
$v_{\rm c,0}$. This was necessary in order to fit the observed 
velocity width distribution of associated low ionization metal absorption. 
Here, we will implement this suppression by reducing the total amount of 
neutral hydrogen in halos below $v_{\rm c,0}$, such that the total mass of 
neutral hydrogen in a halo of mass $M_{\rm v}$ is
\begin{equation}
M_\hi = f_{\rm e} f_\ro{H} \exp \left( -
 \left(\frac{v_\ro{c,0}}{v_{\ro{c}}} \right)^{\alpha_{\rm e}} \right)
 M_{\rm v}
\end{equation}
(see \citetalias{2009MNRAS.397..511B} for a detailed discussion).
We use the fiducial parameters $\alpha_{\rm e}= 3$, $v_{\rm c,0} = 50 \kmsec$.

For the radial distribution of the gas, we assume an 
NFW profile at $z=3$, characterised by a scale radius $r_\ro{s}$ defined in 
\citet{1996ApJ...462..563N}. Following the simulations of
\citet[][Equation (9)]{2004MNRAS.355..694M}, we alter the NFW profile to
give the halo a thermal core at $\simeq 3 r_\ro{s}/4$. The profile is then 
specified by the total mass of the halo $M_{\rm v}$ and
the concentration parameter $c_{\rm v} \equiv r_{\rm v}/r_{\rm s}$. For dependence of the 
concentration parameter on the mass, we 
take the mean value of the $c_{\rm v}-M_{\rm v}$
correlation as given by \citet{2007MNRAS.378...55M},
\begin{equation} c_{\rm v} = c_0 \left( \frac{M_{\rm v}}{10^{11} \Msol}
\right)^{-0.109} \left( \frac{1+z}{4} \right)^{-1} .
\end{equation} 
For dark matter, \citet{2007MNRAS.378...55M} found that $c_0 \approx 3.5$, with 
a log-normal distribution and a scatter around this mean value of 
$\Delta (\ln c_{\rm v}) = 0.33$, in agreement of the results of
\citet{2001MNRAS.321..559B} and \citet{2002ApJ...568...52W}. 
As we will find later, a significantly larger $c_0$ is appropriate for the baryons; 
we will use the column density 
distribution of DLAs to constrain $c_0$ in Section \ref{fNX}.
As we discuss further in Section \ref{limits}, the gas in the
DLAs / faint emitters can be expected to self-shield 
against the meta-galactic ionizing UV background 
at $z\sim 3$. The corresponding self-shielding radius in the 
DM halos we are studying here is generally smaller than the virial radius. 
We therefore set the outer radius of the \hi to be the virial radius
in our modelling and also ignore radiative transfer through the IGM. 
We set $T = 10^4$ K as a fiducial temperature. 

\subsubsection{The assumed kinematics of neutral hydrogen}
The biggest uncertainty in our modelling is probably the 
kinematical state of the gas.
We follow \citetalias{2006ApJ...649...14D} and for our fiducial 
model we assume the gas to be infalling with a 
power-law radial velocity profile\footnote{We use the modification to this law for $\alpha <
0$ given in Equation 10 of \citetalias{2006ApJ...649...14D}.} 
parameterised by $v_\ro{amp}$ and $\alpha$
\begin{equation}
\vv_\ro{bulk}(\rv) = -v_\ro{amp} \left( \frac{r}{r_{\rm v}} \right)^{\alpha} ~ \hat{\rv}
\end{equation}
where $r_{\rm v}$ is the virial radius, for which we follow
the definition of \citet{2004MNRAS.355..694M}. Values in the range $\alpha \in
[-0.5,1]$ should be reasonable. The upper limit describes a spherical
top-hat collapse, while the lower limit represents the accretion of
massless shells onto a point mass ($v^2 \sim GM/r$). 
A spherical top-hat would have $v_\ro{amp} = v_{\rm c}$.

While gas infall will certainly be an important feature of 
the gas kinematics, star formation driven outflows are also likely to
play a role \citep[see][for a review]{2005ARA&A..43..769V}. Note 
that for a spherically symmetric gas
distribution, the red and blue side of the line profiles will just 
be interchanged if the gas is assumed to be outflowing instead of
inflowing with the same velocity profile (ignoring the very slight effects of 
recoil and deuterium). For the more massive and
more actively star-forming LBGs, galactic winds have been suggested to sweep up an 
expanding shell
\citep{2000ApJ...528...96P,2002ApJ...569..742P,2006A&A...460..397V,
2008A&A...480..369S,2008A&A...491...89V,2009MNRAS.398.1263Q}. 
We investigate such a configuration in Section \ref{sect:shells}. 

\begin{figure} \centering 
 	\includegraphics[width=0.4\textwidth]{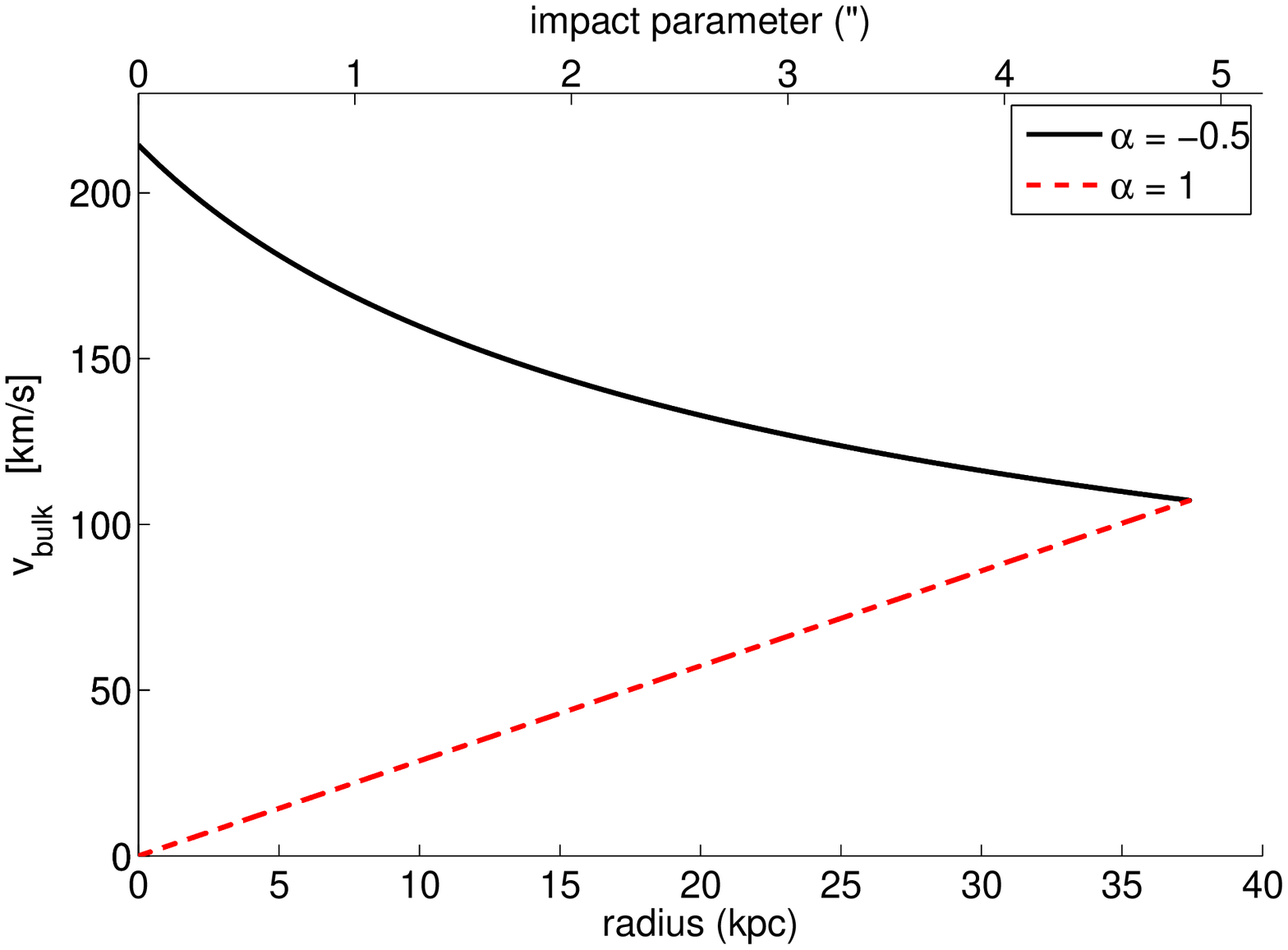}
	\includegraphics[width=0.4\textwidth]{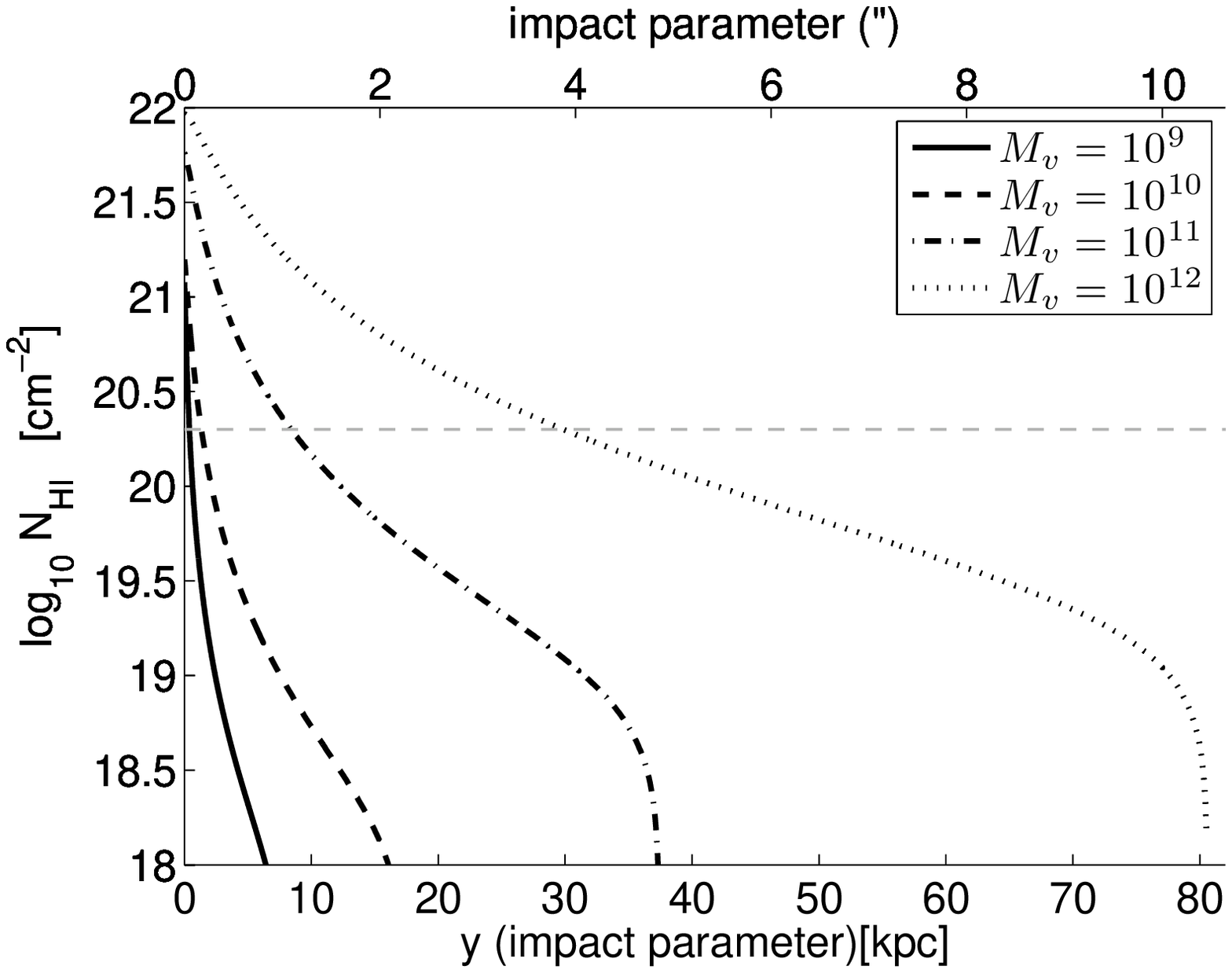}
	\caption{\emph{Top}: The velocity profile for the fiducial
halo, for differing values of $\alpha$ as shown in the legend.
\emph{Bottom}: The column density profiles for the fiducial model with
the total mass as given in the legend in units of
$\Msol$. The column density is calculated along a sightline that passes all the way
through the halo at a distance $r$ from its centre. Note that the $10^9 \Msol$ 
line has been boosted by a factor of $10^4$ to make it visible on the given axes.
The horizontal dashed line indicates the minimum column density of a
DLA.}
	\label{fig:Nrvr}
\end{figure}
 
\subsubsection{The fiducial model} \label{fidmod}
\citetalias{2008ApJ...681..856R} and \citetalias{2009MNRAS.397..511B}
identified star-formation as the most likely source for the 
faint emitters. \citet{2006ApJ...652..981W} used continuum emission
to place stringent limits on \emph{extended} star formation in DLAs. 
We have thus assumed a centrally-peaked emissivity --- all photons 
are created at $\rv = \boldsymbol{0}$. 

Before we attempt to model the data of \citetalias{2008ApJ...681..856R} in detail, 
we will consider the effect on the spectra and surface brightness distribution of
altering the parameters of our model. Our fiducial model parameters
are $(z,M_{\rm v},c_0,f_{\rm e},v_\ro{amp},T) = (3,10^{11} \Msol, 25.3,0.2,
v_\ro{c},10^4 \ro{K})$. The values of $c_0$ and
$f_{\rm e}$ that we have chosen will be justified in Section \ref{fNX}.
The surface brightness $S$ scales with the total luminosity $L_{\lya}$,
which will be given in the caption to each figure. The values of the luminosity chosen 
will be justified in section \ref{sizelum}. The velocity and column
density profiles for the fiducial model are shown in Figure
\ref{fig:Nrvr}, where the column density is as seen along a sightline
that passes all the way through the halo at an impact parameter $y$.

\subsection{\lya radiative transfer in individual halos with gas infall}
In this section, we will consider the effects of changing the most important 
parameters in the model: mass, concentration, velocity profile
(parameterized by the power law index $\alpha$) and baryon 
fraction. Further parameters (temperature and amplitude of the bulk
motions of the gas) are discussed in Appendix \ref{A:Tvampz}.

\subsubsection{The effect of halo mass and concentration}
\begin{figure*} \centering
	\begin{minipage}[c]{0.74\textwidth} \centering 
 		\includegraphics[width=\textwidth]{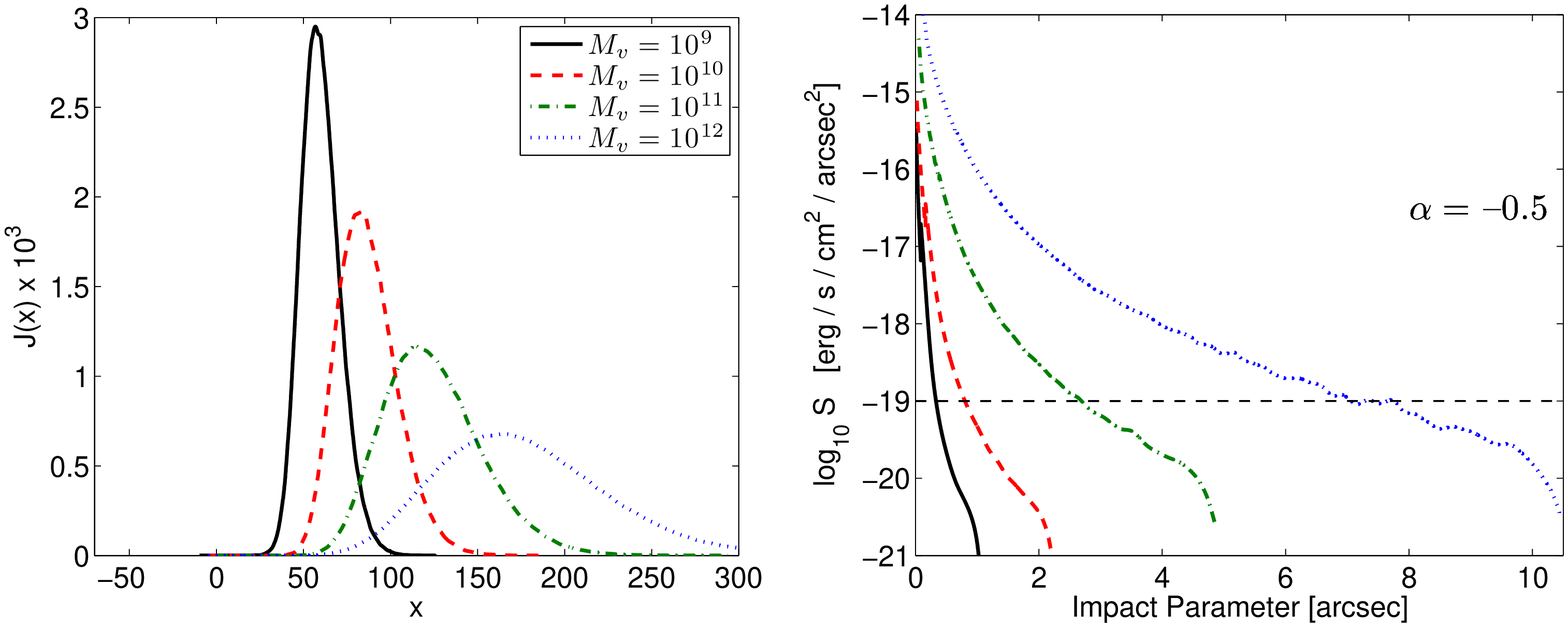}
		\includegraphics[width=\textwidth]{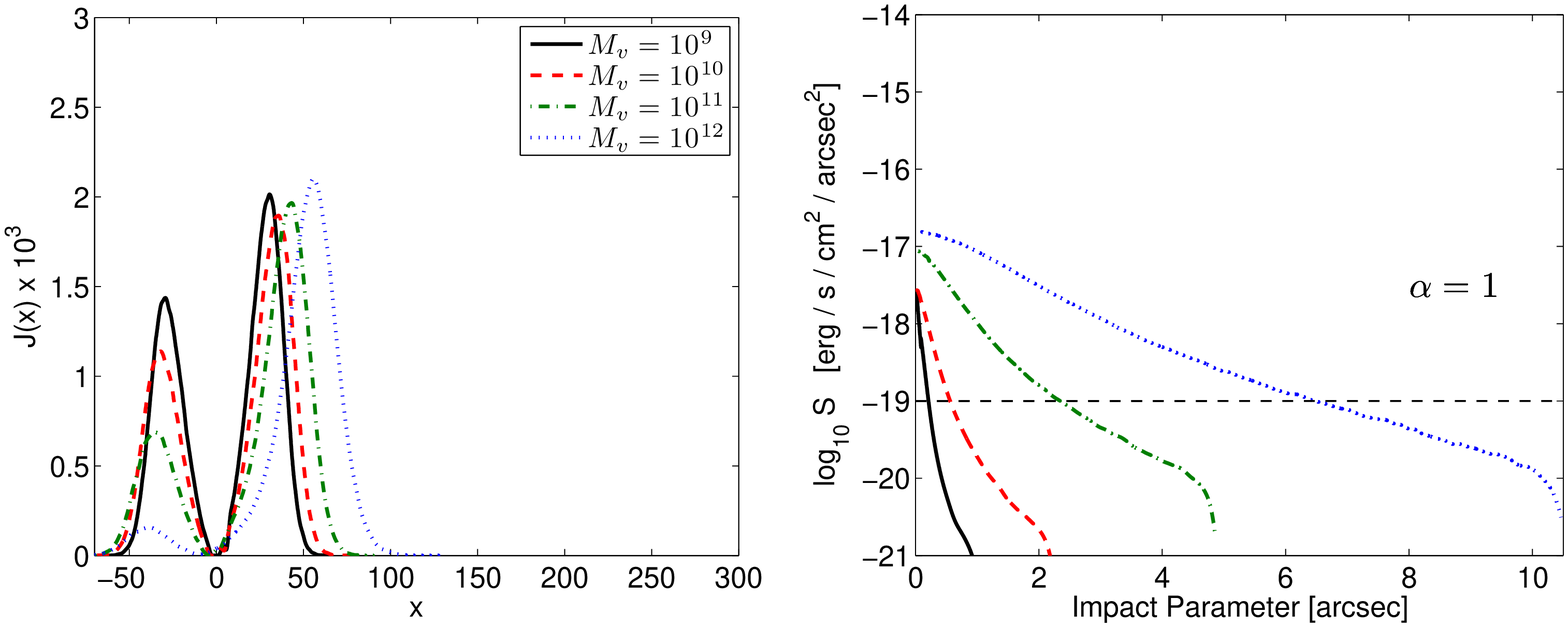}
	\end{minipage}
	\begin{minipage}[c]{0.25\textwidth} \centering 
 		\caption{Spectra (left panels) and surface brightness profiles (right panels) for
emission in halos with total mass as given in the legend in units of
\Msol. The top panels are for $\alpha = -0.5$, with a total \lya luminosity 
which scales with the mass of the DM halo as required to fit the observed
luminosity function in Fig. 10 (see section 4 for more details). 
In the order given in the legend the luminosities are 
$L_{\lya} = (9.6 \times 10^{-6}, 0.88,22,240)\times 10^{42} \ergs$.
The bottom panels are for $\alpha = 1$, with 
$L_{\lya} = (4.8 \times 10^{-7},0.044,1.1,12.1)\times 10^{42} \ergs$.
\emph{Note} that the surface brightness for the $M = 10^9 \Msol$ model
has been raised by a factor $10^4$ to be able to plot it in the same
plotting window. As the mass increases, the emerging photons emerge
bluer, and are scattered to larger radii in the larger halos.
The dashed horizontal line is the detection threshold 
of the Rauch et al. emitters. }
		\label{fig:FbMass}
	\end{minipage}
\end{figure*}

The models of \citetalias{2009MNRAS.397..511B} give us 
a handle on the masses of the halos that host
DLAs. Figure \ref{fig:FbMass} considers halos with masses of $M_{\rm v} = 10^9, 10^{10},
10^{11}, 10^{12} \Msol$, which at $z = 3$ correspond to $v_{\rm c} = 23,
50, 107, 231 \kmsec$, $r_{\rm v} = 8, 17, 37, 81$ kpc, $b_{\ro{max}} = 1.05,
2.25, 4.9, 10.5$ arcsec, where $b_{\ro{max}}$ is the angular radius
corresponding to $r_{\rm v}$.

We see that, as the mass increases, the emerging photons emerge
bluer. This is because, as we add more gas, the central column
densities increase and the photons must shift
further from line centre in order to escape. The $\alpha = 1$ profiles
are increasingly double-peaked for lower masses, while the $\alpha = -0.5$ 
profiles only ever
have one, blue peak. This is because the innermost region of the
$\alpha = 1$ halo has the smallest bulk velocity, and thus most resembles
the uniform, static sphere of Figure \ref{fig:plottest}. As the
bulk velocities increase at the centre of the halo, the amount of energy 
transferred between the gas and photons in each scattering is increased,
favouring one of the two peaks (the blue/red peak for inflowing/outflowing
gas respectively).
 
The surface brightness profile shows that the dominant effect in
increasing the mass is that the virial radius (which we have assumed
to be the outer radius of the \hi) increases. The $\alpha = -0.5$
profile is much more centrally peaked than the $\alpha = 1$ profile. 
The reason is that the larger bulk velocities at the centre of the 
$\alpha = -0.5$ halo can shift the photon into the wings of the
spectral line, resulting in reduced spatial diffusion. 

\begin{figure*} \centering
	\begin{minipage}[c]{0.74\textwidth} \centering
 		\includegraphics[width=\textwidth]{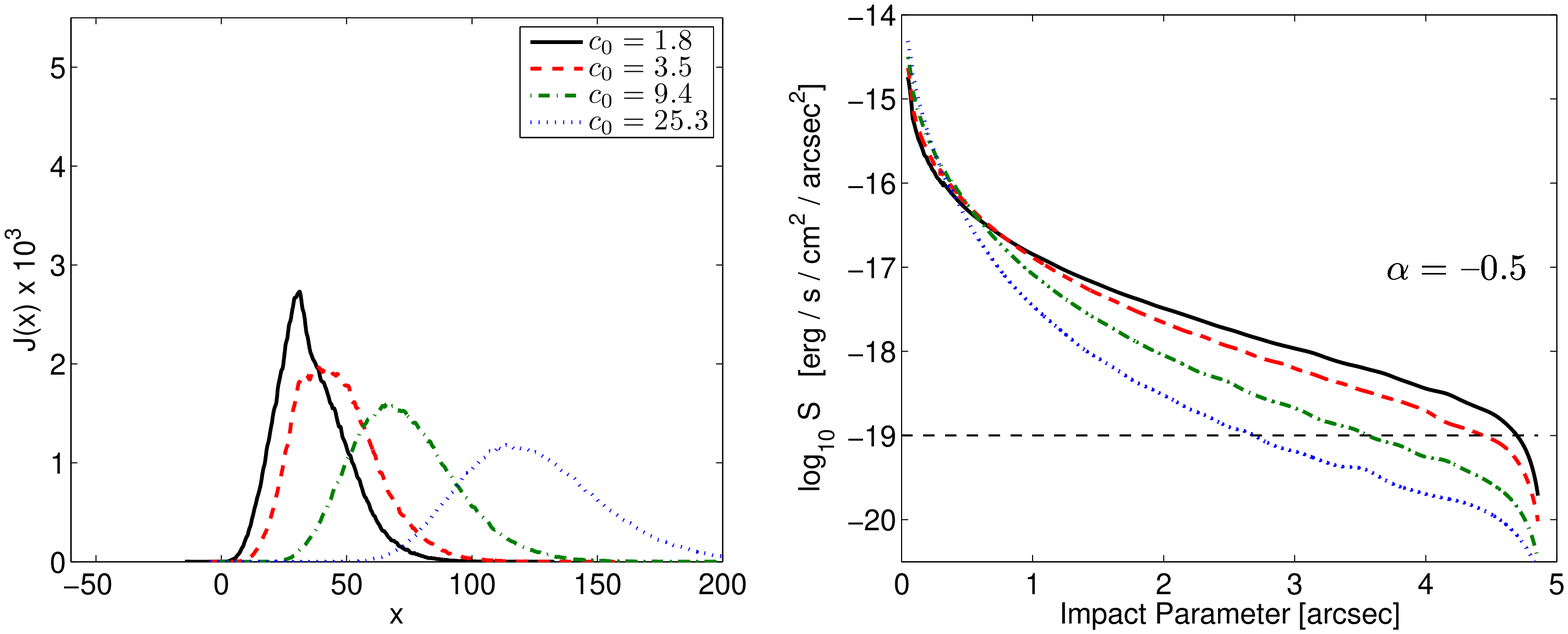}
		\includegraphics[width=\textwidth]{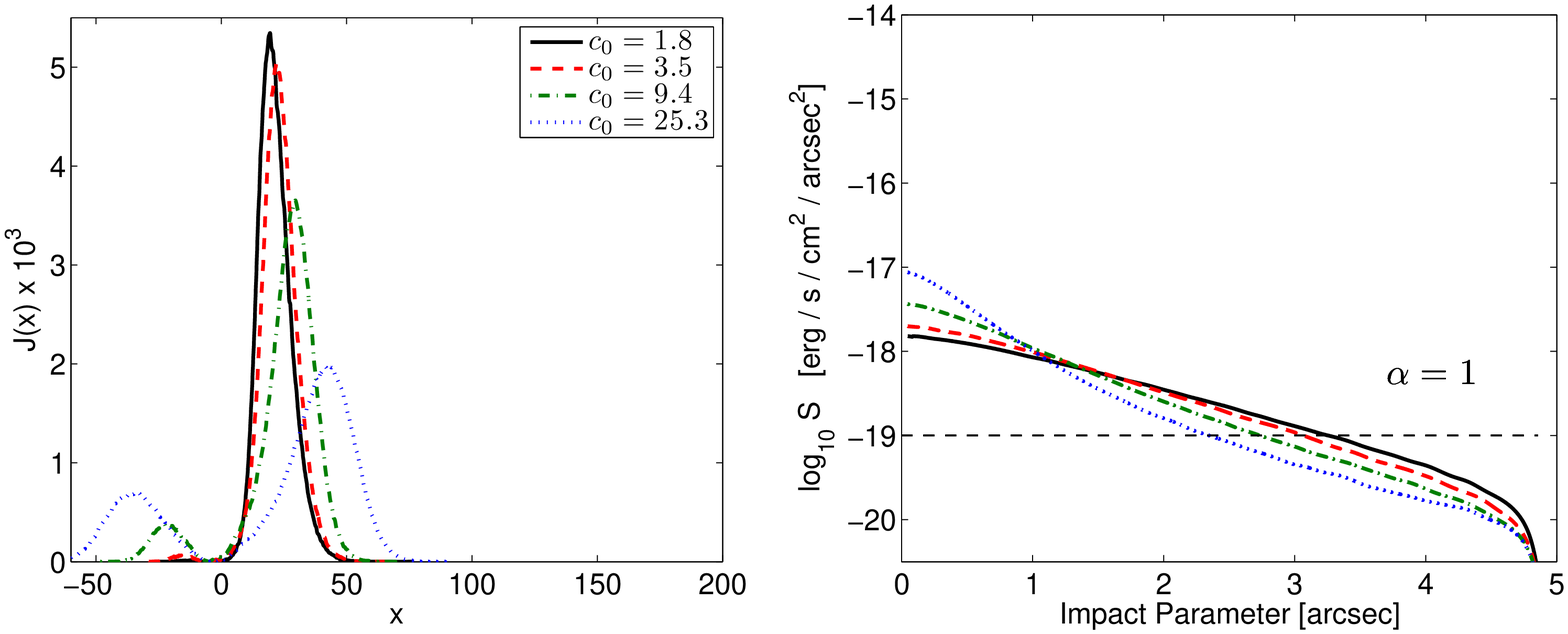}
	\end{minipage}
	\begin{minipage}[c]{0.25\textwidth} \centering 
 		\caption{Spectra and surface brightness profiles for
emission in halos with concentration parameter $c_0$ as given in the
legend. The top panels are for $\alpha = -0.5$, for which 
$L_{\lya} = 2.2 \times 10^{43} \ergs$. The bottom panels are
for $\alpha = 1$, for which $L_{\lya} = 1.1 \times 10^{42} \ergs$.
As the baryon distribution becomes more centrally 
concentrated, the photons emerge bluer. The photons are scattered at 
larger radii in the less concentrated halos. 
The dashed horizontal line is the detection threshold 
of the Rauch et al. emitters.}
		\label{fig:FbCv}
	\end{minipage}
\end{figure*}

For the concentration parameter, we have considered the values 
$1.8,3.5,9.4$ and $25.3$. The spectra and surface brightness profiles 
for these models are shown in Figure \ref{fig:FbCv}.

As the concentration increases, the photons generally emerge
bluer. The $\alpha = 1$ profile becomes \emph{more} double-peaked as the
concentration increases, because the \hi column density
increases at smaller radii, where the bulk velocity is lower. The surface
brightness profile is also more centrally peaked for higher concentrations,
as more scatterings occur at smaller radii. 
 
\begin{figure*} \centering
	\begin{minipage}[c]{0.74\textwidth} \centering 
 		\includegraphics[width=\textwidth]{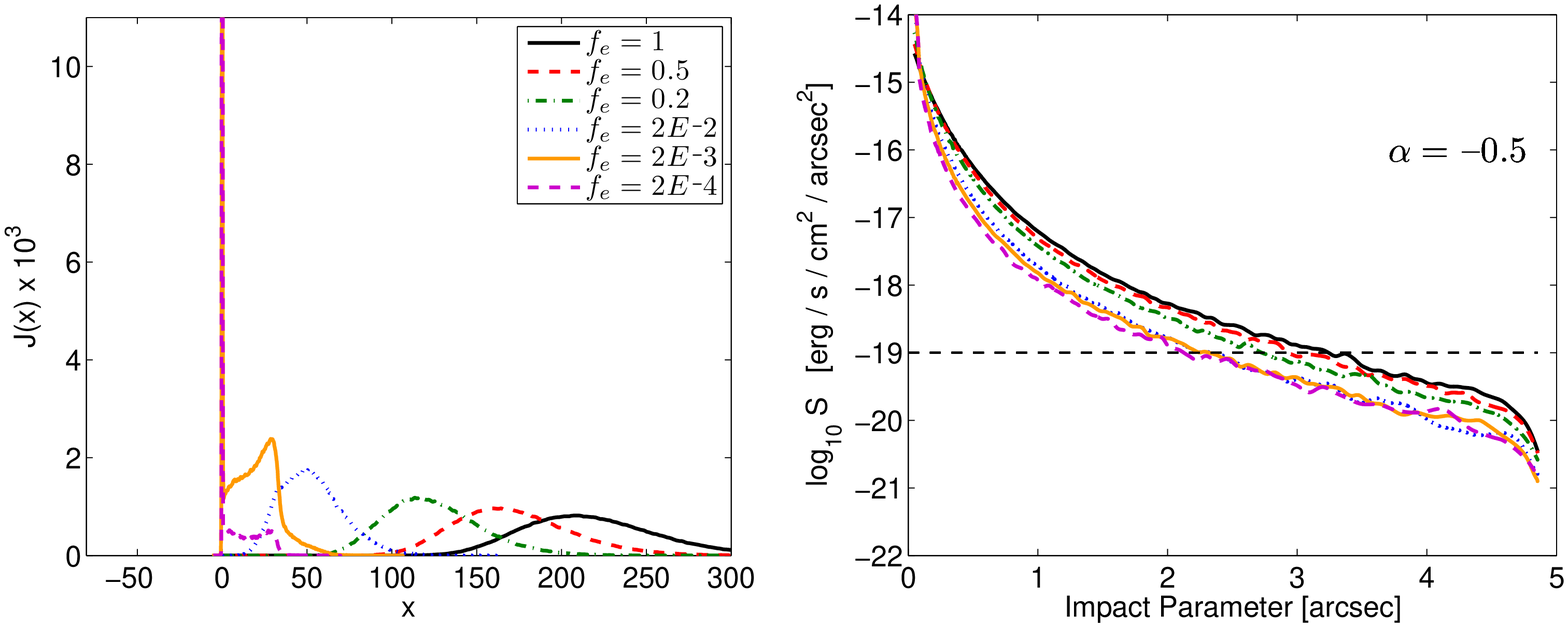}
		\includegraphics[width=\textwidth]{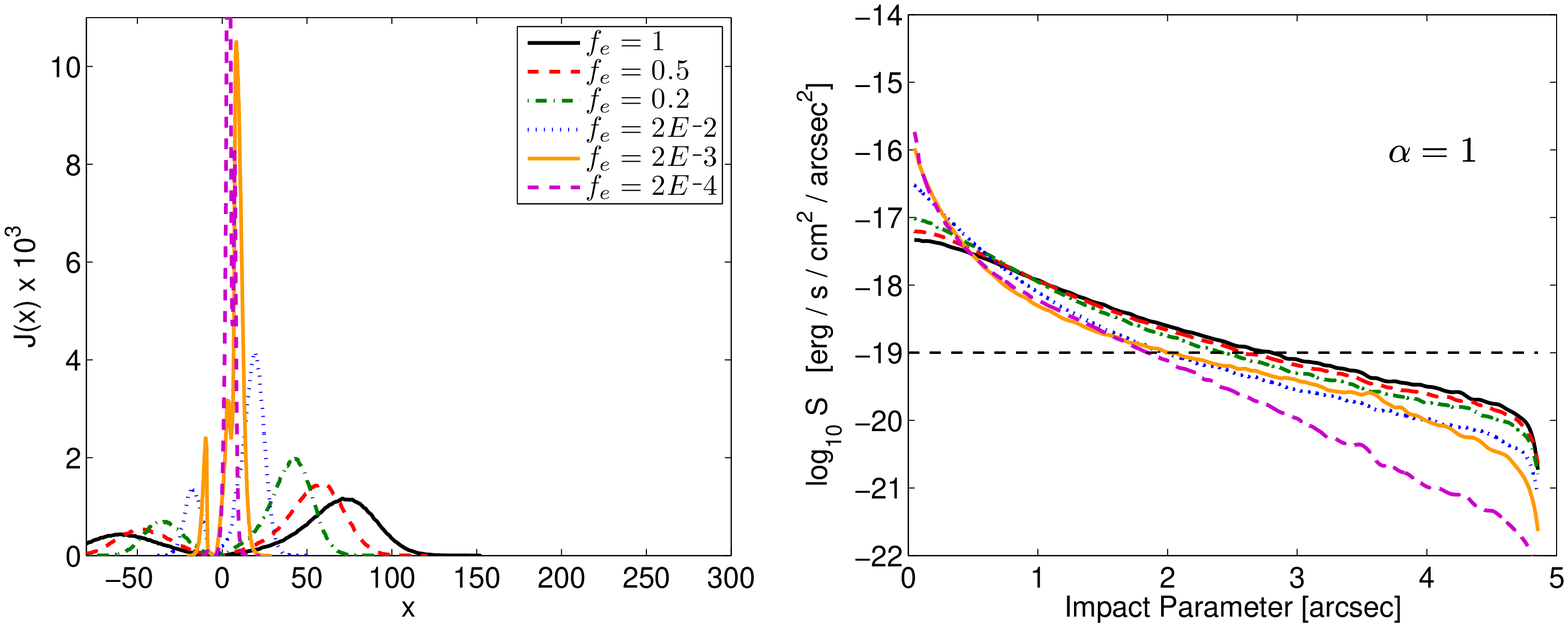}
	\end{minipage}
	\begin{minipage}[c]{0.25\textwidth} \centering 
 	\caption{Spectra and surface brightness profiles for
emission in halos with baryon fraction $f_{\rm e}$ as given in the legend,
where $f_{\rm e} = 1$ corresponds to the cosmic value of the baryon
fraction. The top panels are for $\alpha = -0.5$, the total luminosity 
is kept the same for each model at $L_{\lya} = 2.2 \times 10^{43} \ergs$. The bottom 
panels are for $\alpha = 1$, with $L_{\lya} = 1.1 \times 10^{42} \ergs$.
As $f_{\rm e}$ decreases, the spectral shift decreases 
and the surface brightness becomes more centrally peaked. 
The dashed horizontal line is the detection threshold 
of the Rauch et al. emitters.}
	\label{fig:Fbfe}
	\end{minipage}
\end{figure*}

\subsubsection{The effect of baryonic fraction/column density}
The effect of changing the baryonic fraction $f_{\rm e}$ is shown in 
Figure \ref{fig:Fbfe}. As baryons are removed
from the halo, the spectrum shifts toward $x = 0$ and the surface
brightness profile becomes more centrally peaked, as there is less gas
in the outer parts of the halo to scatter the photons. The baryon
fraction where the photons are not scattered efficiently anymore to
the virial radius (where we have the gas distribution assumed to cut-off)
corresponds to a HI column density\footnote{Which is somewhat dependent on
the spatial profile of neutral hydrogen and the velocity field.} of about $10^{16} \cm$.
For very small $f_{\rm e}$ and $\alpha =
-0.5$, some of the photons can 
escape the halo without scattering at all, 
creating a very narrow peak at $x = 0$. Some of the spectra also show a
trough at $x \approx 6$ due to deuterium.
Note, however, that for $f_{\rm e} \ll 0.1$ the gas in the halos would
not be able to self-shield anymore against the meta-galactic UV
background at $z\sim 3$ so the model would be 
internally inconsistent for such small baryonic fractions.

\subsection{Expanding Shells} \label{sect:shells}

\begin{figure*} \centering
	\begin{minipage}[c]{0.72\textwidth} \centering 
 		\includegraphics[width=\textwidth]{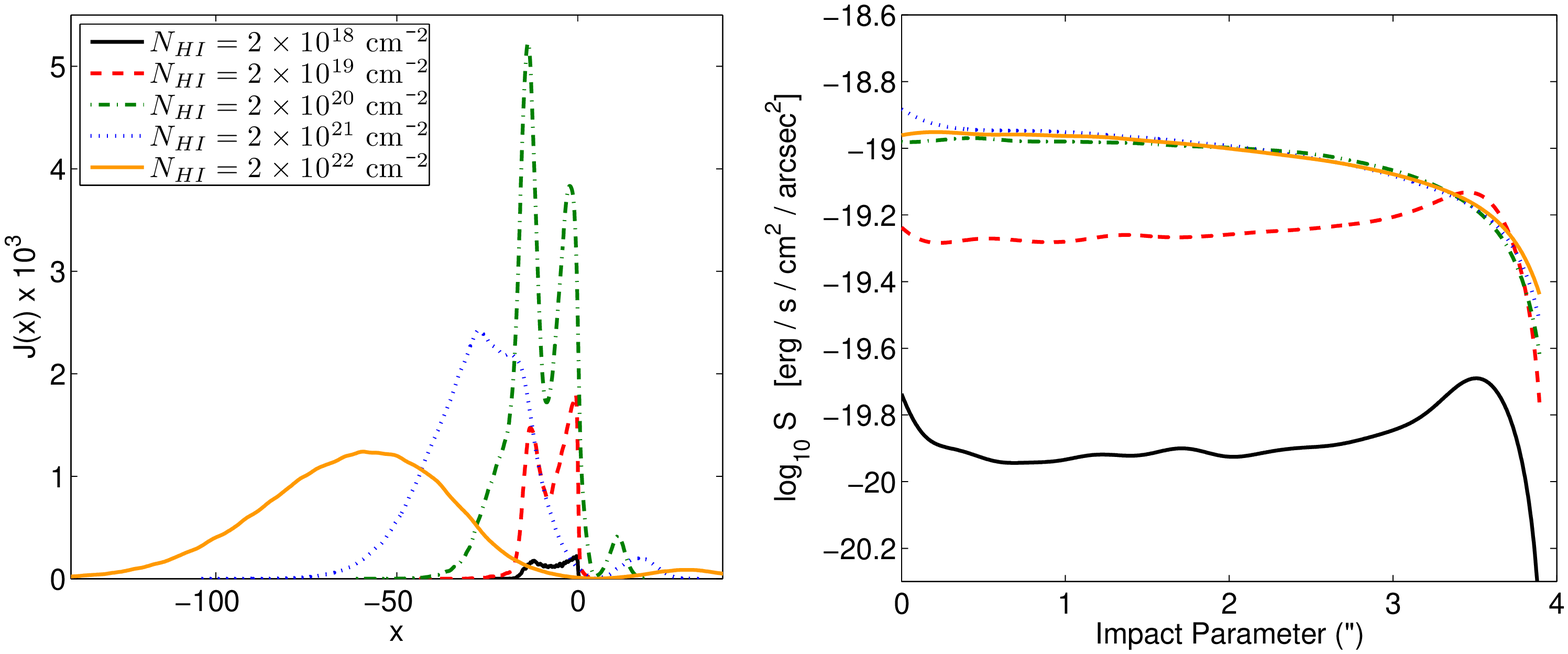}
		\includegraphics[width=\textwidth]{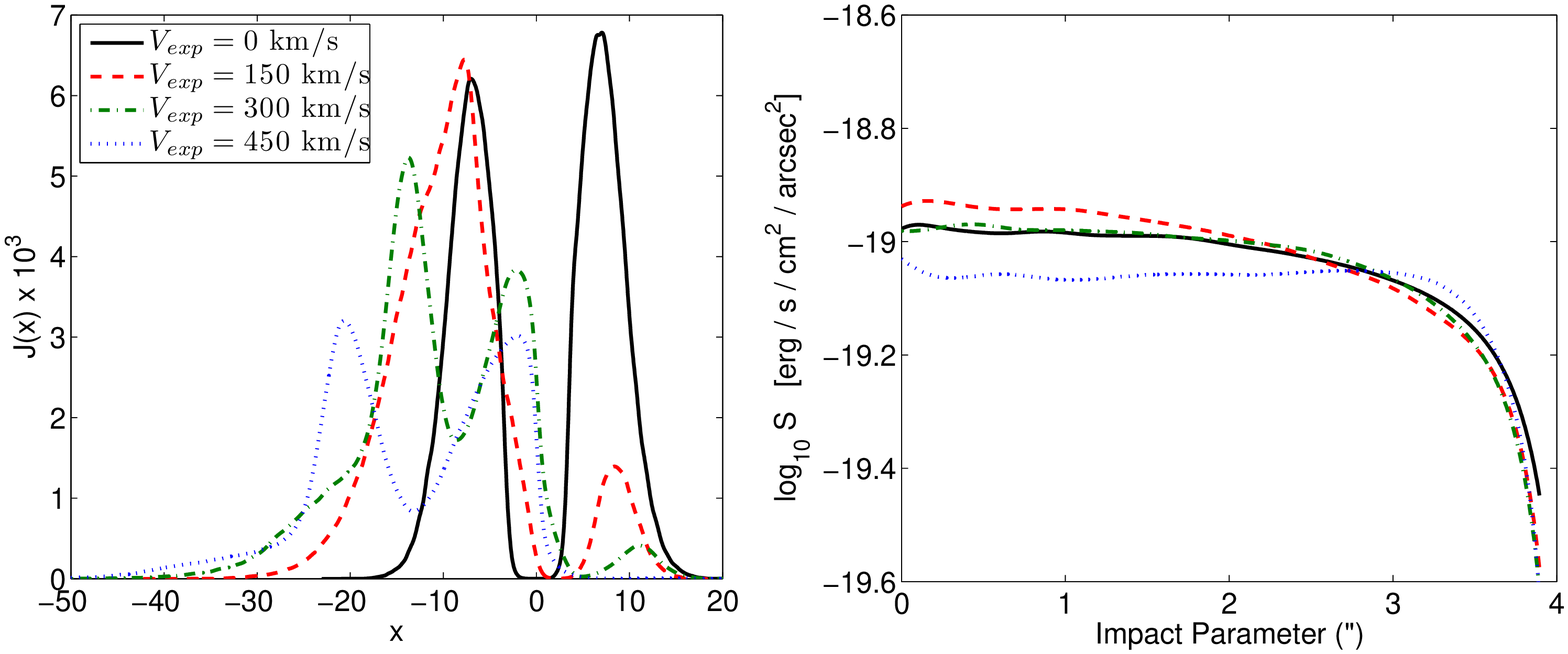}
	\end{minipage}
	\begin{minipage}[c]{0.27\textwidth} \centering 
 		\caption{Spectra and surface brightness profiles for
the modelling of expanding shells. The panels show the effect of
altering the parameters that Verhamme et al. found to be most important, the
column density and the expansion velocity, as given in the
legend. Note that, for small column densities (in the upper panels), a
certain fraction of photons were able to escape directly without
scattering. These photons form a delta function at $x = 0$, and at
small impact parameter, which has been removed from these plots for
clarity. As the column density increases (top panels), the spectrum
becomes redder, and the small blue peak disappears. The most
significant change in the surface brightness profile comes at small
column densities as photons are able to escape directly, leaving less
to be scattered at large radii. As the expansion velocity increases
(bottom panels), the back-scattering mechanism becomes more
pronounced, creating the two red peaks. At small velocities, the 
results become similar to that for a static sphere.}
		\label{fig:Fbsh}
	\end{minipage}
\end{figure*}

For the much brighter \lya emission seen in many LBGs, the \lya emission is 
systematically red-shifted by several hundred \kmsec. This is generally 
attributed to backscattering of the \lya radiation from a wind-driven
expanding shell. \citet{2006A&A...460..397V} used a \lya RT code similar to ours
to calculate the emergent spectrum from an expanding shell. 
We will investigate here in a similar fashion if this picture could apply to the 
\citetalias{2008ApJ...681..856R} emitters. 
We consider a shell of \hi, where the inner radius is a fraction $f_r$
times the outer radius $r_{\rm max}$. The shell is expanding at a uniform
velocity $V_{\rm exp}$, and has a column density of \nhi. The temperature
is set by the Doppler velocity\footnote{This is defined in the same way as the 
thermal velocity dispersion (in the Appendix), with a possible contribution
from a turbulent velocity dispersion, added in quadrature.} $b_T$. Our fiducial model has the
parameters: $(r_{\rm max}, f_r, V_{\rm exp},\nhi, b_T) = (30 \ro{kpc}, 0.9,
300 \kmsec,2 \times 10^{20} \cm, 40 \kmsec)$, similarly to the
fiducial model of Verhamme et al. For the outer
radius of the shell $r_{\rm max}$ we have a chosen a
value similar to the models of the previous sections.

Figure \ref{fig:Fbsh} shows the effect of
altering the parameters that Verhamme et al. found to be most important, the
column density and the expansion velocity. Note that, for small column
densities (upper panels), a certain fraction of photons were
able to escape directly without scattering. These photons form a very sharp
peak at $x = 0$, and at small impact parameter, which has been
removed from these plots for clarity.

As the column density increases (top panels), the spectrum becomes
redder, and the small blue peak disappears, as the photons must
scatter further from line centre in order to escape. The most
significant change in the surface brightness profile occurs for small
column densities, where some of the photons are able to escape
directly, leaving fewer to scatter at large radii.

As the expansion velocity increases (bottom panels), the
back-scattering mechanism becomes more pronounced. Photons which
scatter off the far side of the shell back through its interior are
far enough from line centre to escape through the front side of the
shell. This mechanism creates the two red peaks (the reddest peak
comes from photons that backscatter more than once). At small
velocities, we approach a profile similar to the static sphere. 

The surface brightness profiles are very flat when compared to the
infall/outflow models of the gas in an NFW halo which we had
considered before. Qualitatively, they appear much flatter than the
profiles of the \citetalias{2008ApJ...681..856R} emitters. Almost all of the 
\citetalias{2008ApJ...681..856R} emitters show a central peak. 
One way to produce a central peak for a shell geometry 
is for the column density to be low enough for photons to be able
to escape directly; the resolution of the instrument then broadens the
delta function at $y = 0$ into a central peak. This would limit the
column density of the shell to $\nhi \lesssim 2 \times 10^{19}
\cm$. We conclude that the \citetalias{2008ApJ...681..856R} emitters
appear unlikely to be halo-scale expanding shells of \hi around a central \lya source.

\subsection{Summary of general trends of the \lya emission}
The radial (column) density distribution and the velocity field
are the physical properties which most strongly affect the spectral 
distribution of the \lya emission in our modelling. 
For a spherical density distribution with neither outflow nor infall, 
the distance between the peaks increases with increasing optical depth 
in the same way as the uniform static slab solution. Infall and outflow 
lead to a suppression of the red/blue peak and an increased shift of
the opposite peak. The suppression increases with increasing
velocity amplitude and optical depth. 

The surface brightness profile depends strongly on the
radial velocity profile. With increasing amplitude of the bulk 
motion, the diffusion in frequency space is accelerated and the 
 emission becomes more centrally peaked. In our model 
with increasing amplitude towards decreasing radius 
($\alpha = -0.5$), the average bulk
motions are larger and the effect is more pronounced. 
In the model with the expanding shell, the photons can travel unimpeded 
until they encounter the shell which leads to a rather flat surface
brightness profile. 

The photons diffuse radially until the column density 
of neutral hydrogen density drops to values around $10^{16} \cm$ 
somewhat dependent on the spatial profile of neutral hydrogen and 
the velocity field. The faint extended \lya emission should thus have a rather sharp 
edge which is defined by the surface inside which the gas is able to 
self-shield and the optical depth rises rapidly.

\section{modelling the Rauch et al. emitters} 

\subsection{Surface brightness profiles and spectral shapes}
We begin with a qualitative summary of the properties of the 
\citetalias{2008ApJ...681..856R} emitters. 
Note, however, that due to the faintness of the sources the spectral 
and spatial profiles are rather noisy, making it difficult to identify 
signatures of inflow (more prominent blue peak) or 
outflow (more prominent red peak)\footnote{Recall also that this 
can be more complicated for the case of an outflowing shell, as there
are then two red peaks.}. 
As discussed in \citetalias{2008ApJ...681..856R},
for 12 of the 27 spectra only a single emission peak is visible 
while six/three of the spectra show a weak secondary blue/red counter-peak. 
The remaining spectra are extended in frequency space without a clear peak
structure. The widths of the spectral peaks ranges from $\sim 250 -
1000 \kmsec$, which corresponds to $\Delta x \sim 20-80$ for gas with
a temperature of $10^{4}$ K. 
The surface brightness profiles are predominantly centrally peaked with wings that often extend 
well beyond the Gaussian core of the PSF. This is particularly true of
the brightest sources, while the fainter sources are more difficult
to characterise due to the noise.

\subsection{Simultaneous modelling of DLA properties} \label{fNX}
Modelling the radial distribution of neutral hydrogen 
in DM halos allows us to calculate the column density distribution
along sightlines that intersect the halos. Comparison with the
observed column distribution of DLAs provides a useful 
constraint in this regard. We will use here the same notation as
\citetalias{2009MNRAS.397..511B}.

The column density distribution is defined such that the number of systems ($\dd^2 \cN$) 
intersected by a random line of sight between absorption distance\footnote{The absorption 
distance is defined by \begin{equation} \dd X \equiv \frac{H_0} {H(z)} (1+z)^2 \df z ~.
\end{equation}.}
$X$ and $X+\dd X$, 
with \hi column density between $N_{\hi}$ and $N_{\hi}+\dd N_{\hi}$ is, 
\begin{equation}
\dd^2 \cN = f(N,X) \df X \df N_{\hi}.
\end{equation}
For our model we calculate this quantity using the \ps formalism. 
We need two ingredients. The first is $n_M(M,X)$, the mass function 
of dark matter halos, as calculated by 
\citet{2002MNRAS.329...61S}. The second ingredient is the column 
density of neutral hydrogen in a given halo (of mass $M$ at
absorption distance $X$), as a function of the (physical) impact
parameter $y$, $N_{\hi}(y|M,X)$. This is calculated from the neutral 
hydrogen density as a function of radius. Given that $N_{\hi}$ is a 
monotonically decreasing (and thus invertible) function of $y$, the region between $N_{\hi}$ and 
$N_{\hi}+\dd N_{\hi}$ is an annulus with cross-sectional area $\dd (\pi y^2)$. Hence, 
we can write the column density distribution as, 
\begin{equation}\label{eq:fexp}
f(N,X) = \frac{c}{H_0} \int n_M(M,X) \left| \frac{\dd (\pi y^2)}{\dd N_{\hi}} (N_{\hi}|M,X) \right| \df M .
\end{equation}

\begin{figure} \centering 
	\includegraphics[width=0.47\textwidth]{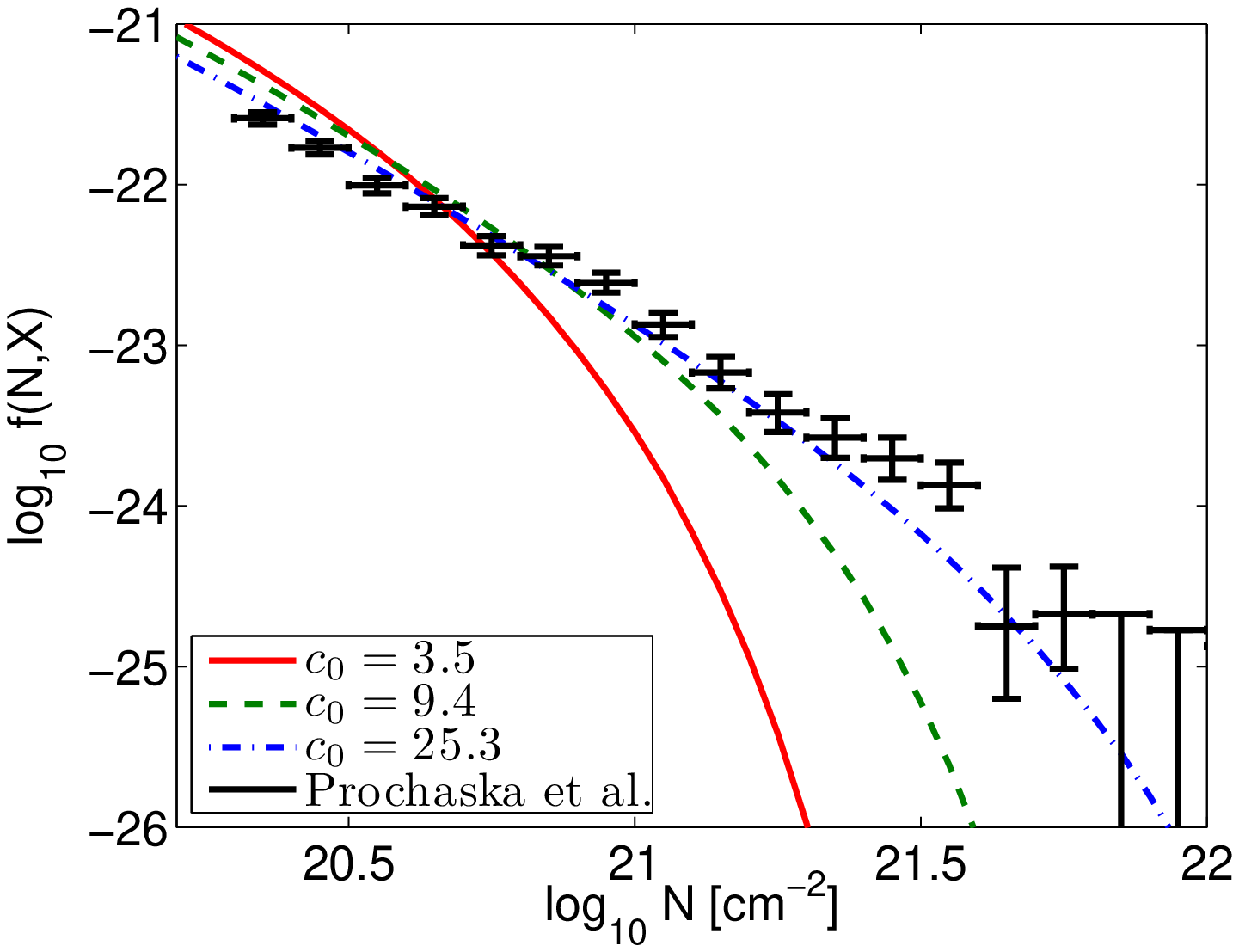}
 	\includegraphics[width=0.47\textwidth]{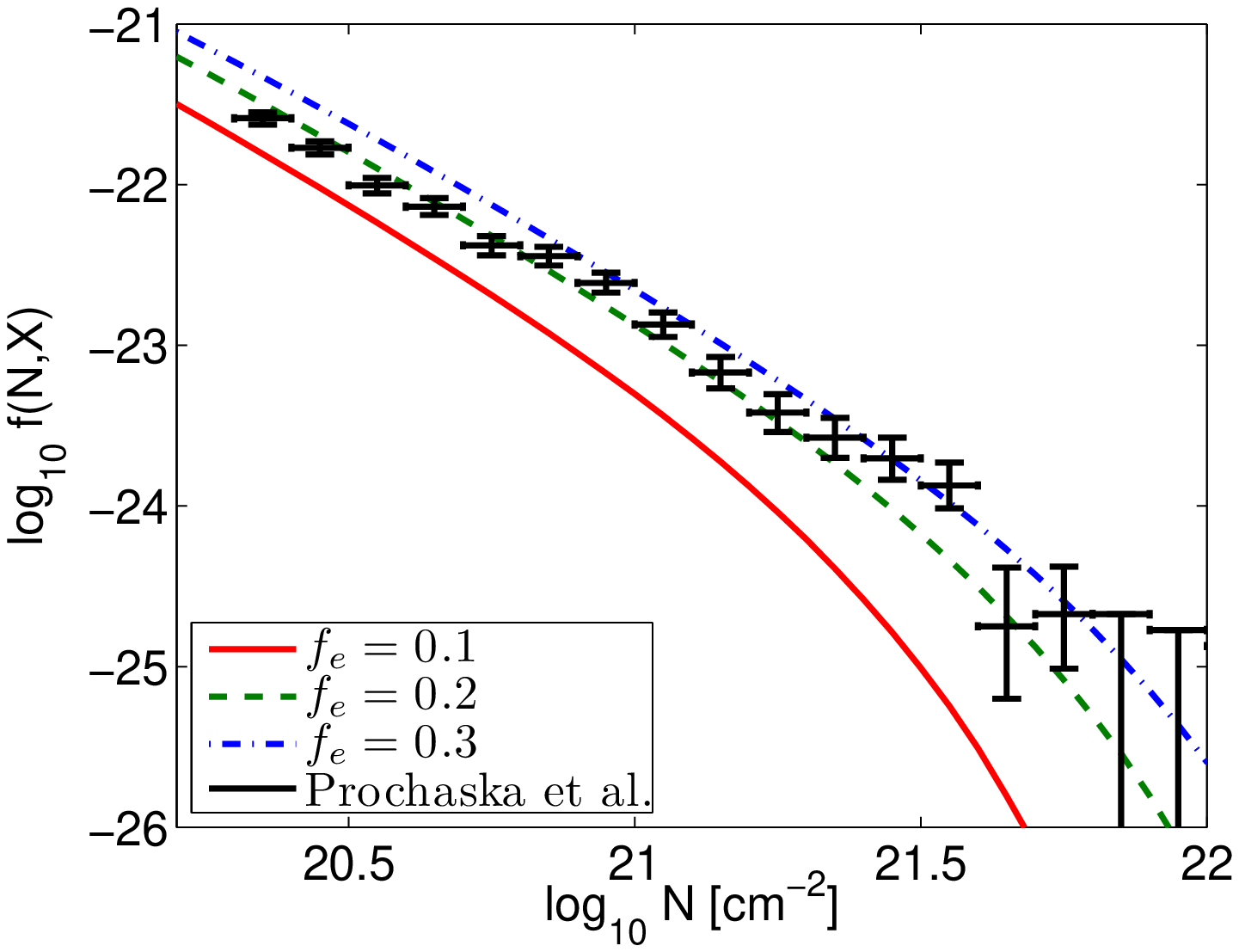}
	\caption{The column density distribution $f(N,X)$ for our model; the black crosses show the 
	data of \citet{2009ApJ...696.1543P}. The upper panel shows the effect of changing the
 concentration parameter of the radial gas density profiles. The lower panel shows the effect 
 of altering the baryon fraction $f_{\rm e}$ (defined relative to the cosmic value).}
	\label{fig:fNX}
\end{figure}

The results are shown in Figure \ref{fig:fNX}. The most important parameters 
(given that we integrate over all M) are 
the normalisation of the concentration parameter for the baryons
$c_0$, and the baryon fraction $f_{\rm e}$ relative to the cosmic value. A good fit to the
observed column density distribution is obtained for 
$c_0 \approx 25.3$ and $f_{\rm e} \approx 0.2$. The figure 
demonstrates the effect of altering these 
two parameters. Decreasing $f_{\rm e}$ decreases the overall normalisation, 
while $c_0$ mostly affects the high column density end of the distribution. 
Changing $v_{\rm c,0}$ has a comparatively small effect. 

Note that our model somewhat overpredicts the number of absorption systems 
with column densities below $N_{\hi} = 20.3 \cm$, in the regime of
Lyman Limit systems (LLs). In this column density range 
the observed $f(N,X)$ flattens significantly \citep{2007ApJ...656..666O}, an
effect which can be attributed to the inability of (super) LLs to
self-shield completely against the meta-galactic 
UV background \citep{2002ApJ...578...33Z}.
Since we have not attempted to model the ionization 
of the gas and in particular the
self-shielding of the gas in any detail, it is not surprising 
that our model does not reproduce this. The UV background should, however, 
not effect our results for the population of predominantly neutral
DLAs. Figure 5 shows that in the LLs regime the inferred size 
of the emission region depends very weakly on column density. 

\begin{figure} \centering 
 	\includegraphics[width=0.47\textwidth]{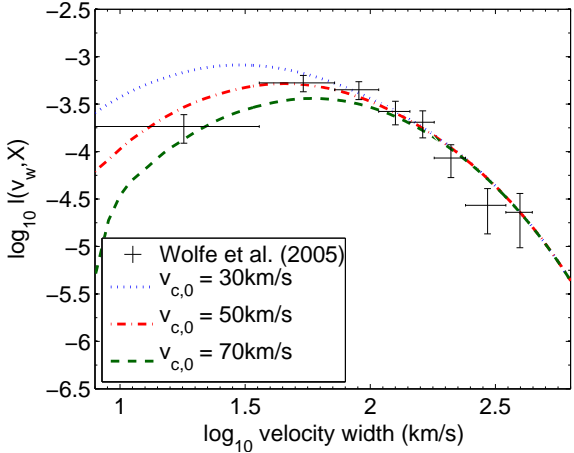}
	\caption{The velocity width distribution $l(v_\ro{w},X)$ of the
associated low-ionization metal absorption of DLAs. The black crosses
show the observational data compiled in Figure 10 of \citet{2005ARA&A..43..861W}.
The legend shows the parameter $v_\ro{c,0}$, below which the baryonic 
fraction is assumed to be suppressed due to the effect of
photo-heating 
and/or galactic winds.}
	\label{fig:vw}
\end{figure}

Similarly to \citetalias{2009MNRAS.397..511B},
we can also calculate the predicted probability distribution of the
velocity width $v_\ro{w}$ of the associated low ionization metal
absorption of DLAs 
for our model as follows,
\begin{equation}
l(v_\ro{w},X) = \frac{c}{H_0} \int_{0}^{\infty} 
	p(v_\ro{w} | v_{\ro{c}}(M)) n_M(M,X) \sigma_{\ro{DLA}}(M,X)
 \dd M, 
\end{equation}
where $p(v_\ro{w} | v_{\ro{c}}(M))$ is the conditional probability distribution 
as discussed in Section 2 of \citetalias{2009MNRAS.397..511B}. The 
DLA cross-section is given by $\sigma_{\ro{DLA}}(M) = \pi y^2_{\ro{DLA}}$, where 
$N_{\hi}(y_{\ro{DLA}}|M,X) = 10^{20.3} \cm$. The result is shown in Figure 
\ref{fig:vw}, along with the observational data of \citet{2005ARA&A..43..861W}

As in \citetalias{2009MNRAS.397..511B} our model fits the data well 
with values of $v_\ro{c,0}$ in the range $50 -70 \kmsec$. 
We should point out that the probability distribution $p(v_\ro{w} |
v_{\ro{c}})$ was originally derived from simulations
which do not include the effect of galactic winds 
and where the distribution of gas in a given halo is 
somewhat different from what we have assumed here
(see \citet{2009MNRAS.397..511B} for a more 
detailed discussion).

We should also emphasize that, rather than using a simple power-law scaling 
for the absorption cross section of DLAs as in
\citetalias{2009MNRAS.397..511B}, we have used here a radial distribution of
neutral hydrogen which is simultaneously consistent with the column density
distribution of DLAs \emph{and} the size distribution of the Rauch et al. 
emitters, as we will see in the next section.


\subsection{ The size distribution and luminosity function} \label{sizelum}
We will not attempt to fit the individual 
rather noisy spectral and surface brightness profiles 
of the \citetalias{2008ApJ...681..856R} emitters here, 
but we will instead focus on the statistical properties 
of the population of emitters.

We calculate the cross-section weighted size distribution $\dd N / \dd z
~ (> r)$ expected from our model as follows. The observations of \citet{2008ApJ...681..856R}
achieved a 1 $\sigma$ surface brightness detection limit of $S_0 = 10^{-19} \ergs$. 
We calculated the expected observed size of our model
emitters by determining the radius $r$ (or equivalently, impact parameter)
at which the surface brightness drops below the \citetalias{2008ApJ...681..856R} limit,
$S(r) = S_0$.

This procedure gives the radius of the emitter as a
function of the mass of the halo. To do this, we need
to specify the intrinsic \lya luminosity $L_{\lya}$ as a function
of mass. Similarly to \citetalias{2009MNRAS.397..511B}, we assume that the 
luminosity is proportional\footnote{The constants of proportionality are chosen 
to that without the suppression below $v_\ro{c,0}$ and with $f_e=0.2$,
$L_0$ defined below has the same value as in
\citetalias{2009MNRAS.397..511B}.} 
to the total mass of neutral hydrogen,
\begin{equation} \label{eq:Llya}
L_{\lya} = L_{0} \left( \frac{M_\hi} {2.4 \times 10^9 \Msol} \right) \ergs.
\end{equation}
This means that the luminosity is subject to the same exponential 
suppression as the (neutral) gas content of DM halos for 
small circular velocities.

\begin{figure}
	\includegraphics[width=0.48\textwidth]{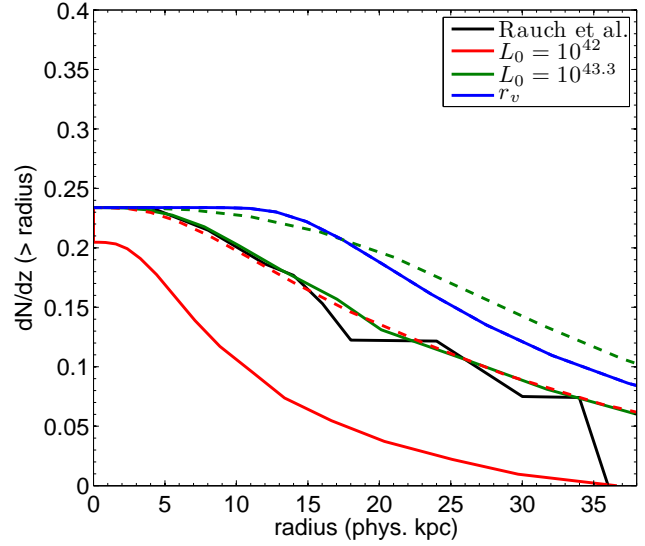}
	\caption{The cumulative size distribution $\dd N / \dd z ~ (> r)$ of the
 \lya emitters, compared with the observations of Rauch et. al (black
 curve). The coloured solid curves are for $\alpha = -0.5$, while the dashed curves are
for $\alpha = 1$. The red and green lines are for the $L \propto M_{\hi}$
model, with the values of $L_0$ as given in the legend. The blue curve assumes
that $r = r_{\rm v}$, that is, assuming we can see
emission all the way to the virial radius. The lines have been
normalised to $\dd N / \dd z = 0.23$ assuming a duty cycle of the \lya
emission with $f_\ro{d}$, such that $n^{\ro{emitters}}_M = f_\ro{d} ~
n^{\ro{halos}}_M$. The values of $f_\ro{d}$ for each model (in the
order they appear in the legend) are $f_\ro{d} = 1, \textbf{0.2},
0.055$ for $\alpha = -0.5$ and $f_\ro{d} = \textbf{0.28}, 0.07, 0.055$ for $\alpha = 1$. 
The models with the two values in bold correspond best to the data.}
	\label{fig:dNdz_r}
\end{figure}

We now calculate the size distribution in the form of the
inferred cumulative incidence rate $\dd N / \dd z ~ (> r)$ and compare it with the data of
 \citetalias{2008ApJ...681..856R} in Figure \ref{fig:dNdz_r}. The
 curves have been normalised to $\dd \cN / \dd z = 0.23$ by assuming 
the emission occurs with a duty cycle $f_\ro{d}$, 
$n^{\ro{emitters}}_M = f_\ro{d} ~ n^{\ro{halos}}_M$. 
The values of
$f_\ro{d}$ for each model are given in the caption to the figure. The
solid curves are for $\alpha = -0.5$, while the dashed curves 
are for $\alpha = 1$. The red and green curves assume $L \propto M_\hi$, with the
values of $L_0$ as given in the legend. The blue curve shows the size distribution 
assuming that $r = r_{\rm v}$, that is, assuming that the emission
is detected all the way to the virial radius.

As we have defined the size of the emitters at a fixed flux level
$S_0$, the observed size of the halos becomes larger as $L_0$ 
increases. For $\alpha = -0.5$ the region with emission 
above this flux level is smaller than in the model with $\alpha = 1$
halos, due to the more centrally concentrated surface brightness
profile. The size distribution flattens at small $r$ due to the
exponential suppression of the luminosity at low $v_{\rm c}$.

Figure \ref{fig:dNdz_r} shows
that we can find values of $L_{0}$ that fit the observed cumulative size
distribution well, where the value
of $L_{0}$ depends on $\alpha$. The required duty cycle is $\sim 20-28 \%$, 
and is fairly insensitive to $\alpha$. The value of $L_{0}$ for 
$\alpha = -0.5$ is rather high, due to the very peaked surface 
brightness profiles in this case.

\begin{figure}
	\includegraphics[width=0.48\textwidth]{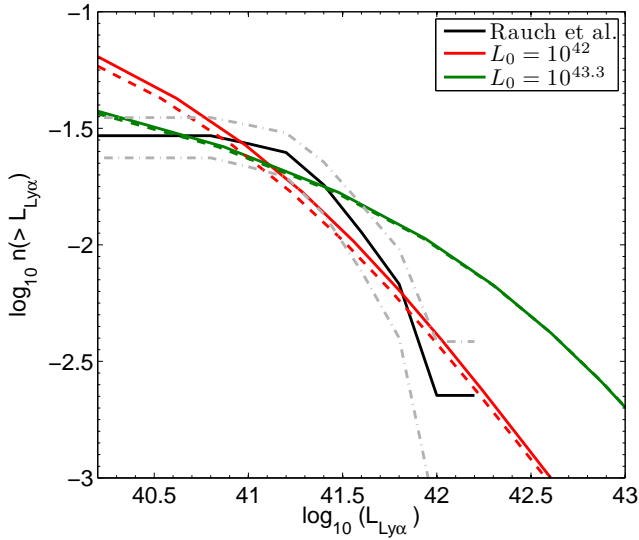}
	\caption{The cumulative luminosity function $n(>L_{\lya})$ of
our model, along with the Rauch et al. data (black solid curve with errors 
indicated by the grey dot-dashed curves). The same models are shown as
in Figure \ref{fig:dNdz_r}, except that the duty cycle is 
chosen to match the luminosity function data. The parameters for each model (in the order
they appear in the legend) are (with negligible dependence on 
$\alpha$): $f_\ro{d} = 0.28, 0.07$. }
	\label{fig:ngm_L}
\end{figure}

Figure \ref{fig:ngm_L} shows the corresponding luminosity distribution
$n(>L_{\lya})$, along with the \citetalias{2008ApJ...681..856R} data
(black solid curve). Note
that the luminosity $L_{\lya}$ predicted by our model is calculated by integrating
the surface brightness inside the radius $r$ where the surface
brightness is above the observational limit. 
Photons that are scattered to radii where the surface brightness falls
below this limit are lost in the noise; the observed luminosity is
thus always less than the intrinsic luminosity. Note further that the 
observed luminosities have not been corrected for ``slit losses''. 
The actual \lya luminosities may thus be a factor two or more larger. 
Note finally, that the intrinsic \lya emission could be
significantly larger due to absorption by dust. 

The figure shows the same models as in Figure
\ref{fig:dNdz_r}, except that the duty cycle is chosen to fit 
the luminosity function data. The solid and dashed curves are 
very close, meaning that there is practically no dependence on 
$\alpha$. The corresponding values of $f_\ro{d}$ 
are given in the caption to the figure. 

The observed lumininosity function is reproduced reasonable well by
our model if we assume $L_0 = 10^{42} \ergs$, the value required to
match the observed size distribution for our model with $\alpha =1$
This model predicts a
somewhat steeper faint end slope than appears to be observed, 
but as discussed in 
\citetalias{2008ApJ...681..856R} it is uncertain whether the turn-over
in the observed luminosity function is real or due to incomplete 
identification of emitters close to the detection threshold. 
Should the turnover consolidate with deeper data then this may suggest 
a somewhat sharper cut-off of the efficiency for \lya emission 
in shallow potential wells as we had e.g. assumed in \citetalias{2009MNRAS.397..511B}. 
The required duty cycle is $f_\ro{d} = 0.28$ and thus comparable
to the value required to fit the observed size distribution. 
With $\alpha=1$, $L_0 = 10^{42} \ergs$, $f_\ro{d} \sim 0.28$ 
our model therefore fits both the observed size and luminosity 
distribution of the \citetalias{2008ApJ...681..856R} emitters. 
As previously discussed, the model with $\alpha=-0.5$ has strongly centrally 
peaked emission. As is apparent 
from Figure \ref{fig:ngm_L} this leads to a mismatch 
with the observed luminosity function if we fix $L_0=10^{43.3}$
to fit the observed size distribution for this value of $\alpha$. 
Even for a significantly smaller duty cycle of $f_\ro{d} = 0.07$ 
we cannot match the shape of the observed luminosity function, which in this case 
is significantly steeper than predicted by the model over the full
range of luminosities. 

We therefore conclude that we can successfully reproduce 
both the absorption properties of DLAs and the \lya emission data of 
\citetalias{2008ApJ...681..856R}, using a self-consistent model 
with centrally concentrated, star-formation
powered \lya production with a duty cycle of $\sim 25 \%$, 
coupled with radiative transfer effects that
set the observed size of the emitters \emph{if} the velocity 
field of the gas facilitates the scattering of the photons to large radii
with moderate central bulk velocities as in the model with $\alpha = 1$. 
Our model's success in simultaneously reproducing the absorption
properties of DLAs significantly strengthens the assertion of \citet{2008ApJ...681..856R} that 
the faint emitters are in fact the host galaxies of DLAs.

\begin{figure}
	\includegraphics[width=0.48\textwidth]{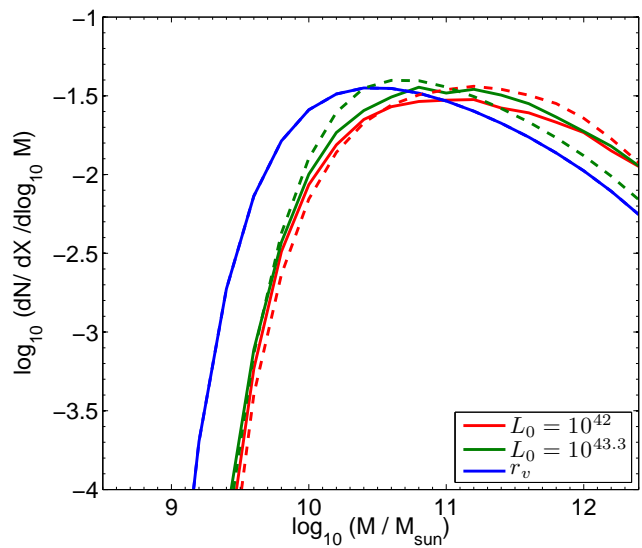}
	\caption{ The contribution of different mass ranges to the
incidence rate of DLAs/emitters ($\dd^2 \cN / \dd X / \dd \log_{10} M$). The
same models are shown as in Figure \ref{fig:dNdz_r}. The majority of
DLAs/emitters have masses in the range $10^{10} - 10^{12} \Msol$. We refer the
reader to \citetalias{2009MNRAS.397..511B} for a comparison of our
prediction of $\dd^2 \cN / \dd X / \dd \log_{10} M$ to that of numerical
simulations. }
	\label{fig:dNdXM}
\end{figure}

Finally we calculate $\dd^2 \cN / \dd X / \dd \log_{10} M$, the 
contribution to the incidence rate from halos of different masses. The 
result is shown in Figure \ref{fig:dNdXM}. The majority of
DLAs/emitters have masses in the range $10^{9.5} - 10^{12} \Msol$ for
all models. This is a similar range of masses to that found in the numerical simulations of 
\citet{2007ApJ...660..945N,2008ApJ...683..149R,2008MNRAS.390.1349P,2009MNRAS.397..411T}. 
We refer the reader to \citetalias{2009MNRAS.397..511B} for a more
detailed comparison of our prediction of $\dd^2 \cN / \dd X / \dd
\log_{10} M$ with that of numerical simulations. The success of \citet{2008MNRAS.390.1349P} 
in reproducing the observed metalicity distribution is additional
evidence in favour of this mass range, as are the observed spatial
correlation of DLAs and LBGs \citep{2006ApJ...652..994C}.

\subsection{Differences to our previous modelling}

In \citetalias{2009MNRAS.397..511B}, we presented
modelling in which we assumed simple scaling
relations for the sizes and luminosities of the Rauch et al. emitters
with the masses/circular velocities of their DM host haloes. 
The scaling relation for the sizes was
motivated by the modelling of the velocity width
distribution of the associated low ionization metal absorption of
DLAS by \citet{1998ApJ...495..647H}.
No attempt to model the spatial and spectral distribution of the
\lya emission was made in \citetalias{2009MNRAS.397..511B}. It is thus 
gratifying that our modeling here with \lya
radiative transfer through inflowing/outflowing distributions 
of neutral hydrogen in the emitters for which the radial profile 
is constrained by the observed column density ditribution is able 
to still fit the luminosity function and size distribution 
of the emitters as well as the velocity width distribution of 
DLAs for a specific choice of the scaling of the inflow/outflow velocity 
with radius. This is a non-trivial result and means that the
radiative transfer calculations as constrained by the observed
properties of the Rauch et al. emitters and DLAs  produce a 
scaling relation of the sizes of emitters with the
masses/circular velocities of their DM host haloes very simular 
to that assumed by \citetalias{2009MNRAS.397..511B}.

\subsection{Limitations of the current modelling} \label{limits}
Our modelling has a number of limitations and (over-) simplifications that we will 
discuss in this section. Firstly, we have assumed spherical symmetry throughout. 
Deviations from a uniform spherical configuration will most likely make escape 
easier for \lya photons in certain directions. 
Similarly, our simple velocity field is certainly an
oversimplification. More complex velocity fields may bring photons 
passing through gas in the outer regions of the halo back toward 
line centre. This may also increase the spatial extent of the emitter.

We have also ignored radiative transfer outside the virial radius of
the halo. This effectively assumes not only that scattering in the
outer parts of DM halos is negligible but that the same is true 
for scattering by the neutral hydrogen contained in the IGM. Whether
these are reasonable assumptions will depend both on the poorly
constrained distribution of neutral hydrogen in the outer parts of low-mass
DM halos and on the bulk motion of the
scattering gas relative to the IGM (see e.g. \citet{2004MNRAS.349.1137S}).

We have furthermore not modelled the effect of ionising radiation, 
either from external or internal sources of ionising photons, except by reducing the
amount of neutral hydrogen within the virial radius of 
the DM halos below its cosmic value,
ostensibly due to the effects of the UV background. More
sophisticated modelling should take into account the self-shielding of
the gas against the meta-galactic UV background
self-consistently. We have checked here that the radius at which the 
gas is able to maintain neutrality by self-shielding is $\gtrsim r_\ro{emission}$, 
the size of the emission region at the flux level of the Rauch et al. emitters 
as defined in Section \ref{sizelum}. The meta-galactic UV background 
should thus have little effect on the size of the \lya emission
region predicted by our modelling. Note, however, that this will not
be the case anymore at fainter flux levels as the neutral hydrogen 
density/column density will drop sharply outside the region 
able to self-shield. 

Internal sources of ionising stars such as stars would reduce the neutral 
fraction of the gas, and in an inhomogeneous manner. Neither stars nor
the UV background are expected to 
significantly ionise the bulk of the neutral gas in the proto-galaxies
studied here --- these are the sites of 
DLAs after all. Spatially extended star formation is likely 
to increase the production of \lya 
at large radii, possibly increasing the observed size of the emitters.

Finally, we have also ignored the effects of dust. Our assumption of
a negligible effect of dust may not be unreasonable as DLAs 
are known to have a rather small dust content
\citep{2005ARA&A..43..861W} and the escape fraction 
for \lya emission appears to increase strongly towards unity with decreasing 
stellar mass of the emitters \citep{2009arXiv0911.2544O}. 
The recent modelling of \lya radiative 
transfer by \citet{2009arXiv0907.2698L} including dust also suggest that for the
majority of DLAs, with their rather low metallicity 
which are hosted by rather low mass halos, the effect of dust is
probably not too important. Note however that there are counter-examples. 
As shown by \citet{2009A&A...502..791A}, 
models with almost complete absorption of dust can explain the fact that 
the expected \lya emission from the star formation in 
IZw 18 appears to be entirely absorbed despite a rather modest
reddening.  As discussed in section 3.1.1, the presence of dust 
may actually be (partially) responsible for the smaller than unity
duty cycle which we require to fit the luminosity function. 
Note also that the effect of dust should depend sensitively 
on the clumpiness of the neutral gas \citep{2006MNRAS.367..979H}.


\section{Conclusions}
We have used a Monte-Carlo radiative transfer code to 
model the spatial and frequency distribution of the \lya emission 
due to star formation in (proto-)galaxies at the centre of DM halos
with masses of $10^{9.5}$ to $10^{12}\Msol$ for a range of assumptions 
for the spatial distribution and the dynamical state of neutral hydrogen.
DM halos in this mass range had been previously identified as the likely hosts 
of DLAs and the recently detected population of faint 
spatially extended \lya emitters.
Our main results are the following:

\begin{itemize}

\item As previously found by other authors, the 
spectral shape of the \lya emission from star formation in 
galaxies for which the dynamics of neutral hydrogen is dominated by infall/outflow
is characterized by a strong blue/red peak, occasionally 
accompanied by a weaker red/blue peak. The spectral shape is very
sensitive to the spatial distribution,
velocity structure and (to a smaller extent) the 
 temperature structure of the gas. 
The larger central column densities in the more massive
galaxies/halos make escape of the \lya photons more difficult. The 
photons must then scatter further in frequency and space and thus
emerge with a larger frequency shift and the emission extends to 
larger radii. Larger bulk motions lead to more energy per scattering 
being transferred between the photons and the gas. This results in a 
larger frequency shift of the dominant spectral peak and a larger
contrast between the strong and weak spectral peak.

\item The surface brightness profiles for photons emitted at the 
centre of the halos show a central peak with wings extending as far as our assumed 
neutral hydrogen distribution as long as the column density of neutral
hydrogen exceeds about $\sim 10^{16} \cm$. 
The spatial profile of the emission 
is likewise sensitive to the spatial distribution, velocity 
and temperature structure of the gas. The spatial distribution is significantly more 
centrally peaked when the amplitude of the bulk motions increase toward the centre of the halo.

\item Expanding shells of neutral hydrogen similar to those invoked 
to explain the \lya emission from LBGs produce spectra with one or more 
prominent red peaks. The surface brightness profiles are very flat, remaining 
essentially constant for 75\% of the radius of the shell. This appears
at odds with the observed profiles of the \citetalias{2008ApJ...681..856R} emitters, 
almost all of which show a central peak.

\item Our modelling simultaneously reproduces the column density
 distribution of the neutral hydrogen and the velocity width 
 of the associated low ionisation metal absorption of DLAs, 
 as well as the size distribution and the luminosity 
 function of the Rauch et al. emitters if we assume 
 i) that absorbers and emitters are hosted by DM halos 
 that retain about 20\% of the
 cosmic baryon fraction in the form of neutral
 hydrogen, with a spatial distribution which follows a NFW
 profile with concentration parameter $\sim 7$ times larger 
 than that of the dark matter,
 ii) that absorbers and emitters are hosted by DM halos 
 with virial velocities $\ga 50 \kmsec$, and
 iii) that the central \lya emission due to star formation 
 has a duty cycle of $\sim 25\% $ and the luminosity is proportional to 
 the mass of neutral hydrogen in the DM halos.

\item The DM halos that contribute most to the incidence rate of DLAs have masses in 
 the range $M \sim 10^{9.5}$ - $10^{12} \Msol$ and virial velocities in
 the range of 35 to 230 \kmsec. The lower cut-off is mainly determined by the 
 rather sharp decrease in the velocity width distribution of the
 associated low-ionization metal absorption in DLAs at velocity
 width $\la 30 \kmsec$ but may also be reflected in the turn-over 
 of the \lya luminosity function at the faintest fluxes. The
 DM host halo masses are significantly 
 smaller than those inferred for $L_*$ LBGs, consistent 
 with the much higher space density of the faint emitters. 
\end{itemize}

The success of our detailed \lya radiative transfer modelling 
in explaining the observed properties of both DLAs and the 
faint Rauch et al emitters with a consistent set of assumptions
further strengthens the suggestion that the faint emitters are 
indeed the long-searched for host galaxies of 
DLA/LLs. Together with our modelling, the observed properties 
of the faint emitters should thus provide robust estimates not only 
of the space density, \lya luminosity and extent of the gas
distribution but also of the masses of the DM halos, the duty 
cycle of star formation and the spatial profile and
kinematics of the gas distribution of a statistically 
representative sample of DLA host galaxies. The 
current ultra-deep spectroscopic surveys in the 
HUDF and HDF should soon provide important additional 
information on the stellar content and possibly also dust content 
of these objects which will allow to further test the nature of what 
are almost certainly the building blocks of typical 
present-day galaxies like our own.


\section*{Acknowledgments} LAB is supported by an Overseas Research
Scholarship and the Cambridge Commonwealth Trust. We thank Richard
Bower and Bob Carswell  for comments and the referee for a helpful
report which has improved the manuscript.


\appendix

\section{\lya Radiative Transfer Algorithm} \label{A:lyaRT} 
\subsection{Structure of the Monte-Carlo Code}
The details of each step in our Monte Carlo code for \lya RT are outlined
below --- throughout, $R_i$ (for $i = 1,2,3,\ldots$) denotes a random
number generated uniformly between 0 and 1.

\begin{enumerate}
\item We begin by specifying, as a function of position, \rv
\begin{itemize}
\item[-]{$\dhi(\rv)$}, the number density of H\textsc{i},
\item[-]{$\epsilon(\rv)$}, the \lya emissivity (in photons/s/cm$^{3}$),
\item[-]{$\vv_\ro{b}(\rv)$}, the bulk velocity field of H\textsc{i}, and
\item[-]{$T(\rv)$}, the temperature of H\textsc{i}.
\end{itemize}
For the modelling in this paper, these quantities are discussed in Section 
\ref{S:gasmodel}.

\item \label{generate} We generate a photon at an initial position
$\rv_i$ according to the emissivity. We then choose the photon's initial
direction \nuv from an isotropic distribution, 
\begin{equation} \nuv = (\sin\theta \cos\phi, \sin\theta \sin\phi,
\cos\theta)
\end{equation} where
\begin{align} &\theta= \cos^{-1}(2R_1 -1)			&\quad
&\ro{(the polar angle)}, \\ &\phi= 2 \pi R_2
&\quad &\ro{(the azimuthal angle).}
\end{align}

We generate the photon's initial frequency as follows. The
emission (and absorption) profile in the rest
frame of the emitting atom is assumed to be 
Lorentzian,
\begin{equation} \phi_{\ro{L}}(\nu) = \frac{\DnuL/2 \pi} {(\nu -
\nu_0)^2 + (\DnuL/2)^2}, 
\end{equation} where $\nu_0 = 2.47 \times 10^{15} \ro{Hz}$ is the
central \lya frequency and $\DnuL = 9.936 \times 10^{7} \ro{Hz}$ is the
natural line width. We take into account the thermal Doppler
broadening with an Maxwell-Boltzmann (MB) distribution of the velocity of the scattering
atoms. The line profile can then be written as an average of the Lorentz profile over
the atoms' velocities,
\begin{align} &\phi(\nu) = \int\limits_{-\infty}^{\infty}
\phi_{\ro{L}}\left( \nu - \frac{\nu_0 v_z}{c} \right)
\frac{1}{v_{\ro{th}} \sqrt{\pi}} \exp \left( -\frac{v_z^2}
{v_{\ro{th}}^2} \right) \df v_z \label{eq:profilenu} \\ \Rightarrow \
&\phi(x) = \frac{1} {\DnuD \sqrt{\pi}} H(a,x), \label{eq:profilex} 
\end{align} where the argument of $\phi_{\ro{L}}$ takes into account
the Doppler shift in the frequency of the photon as seen by an atom
with velocity component $v_z$ in the direction of the photon's
motion. The other quantities are
\begin{align} &v_{\ro{th}} = \left( \frac{2 k T}{m} \right)^{1/2}
= 12.85 \kmsec \left( \frac{T} {10^4 \ro{K}} \right)^{\frac{1}{2}}, \\
&a = \frac{\DnuL} {2\DnuD} = 4.693 \times 10^{-4} \left( \frac{T}
{10^4 \ro{K}} \right)^{-\frac{1}{2}}, \\ &\DnuD = \left(
\frac{v_{\ro{th}}}{c} \right) \nu_0 ,\\ &x = \frac{\nu - \nu_0}
{\DnuD}, \end{align} the thermal velocity dispersion, 
relative line width, Doppler frequency width and frequency displacement in
units of $\DnuD$, respectively. We will use $x$ as our
frequency variable. We have also made use of the \emph{Voigt
function},
\begin{equation} \label{eq:voigt} H(a,x) \equiv \frac{a} {\pi}
\int\limits_{-\infty}^{\infty} \frac{e^{-y^2}}{(x-y)^2 + a^2} \df y.
\end{equation} The Voigt function can be approximated as a Doppler core and Lorentz
wings,
\begin{equation} \label{eq:voigtapp} H(a,x) \sim \begin{cases}
e^{-x^2},						& |x| < x_c \quad\text{(core),} 
\\ \frac{a}{\sqrt{\pi} x^2}, 	& |x| > x_c \quad\text{(wing)} .
\end{cases}
\end{equation} The transition between the two occurs at a `critical'
frequency $x_c$, defined as the solution to $e^{-x_c^2} = a/\sqrt{\pi}
x_c^2$. For $T = 10^4$ K, we have $x_c \approx 3.255$. We will use 
the approximation to the Voigt function given in \citet{2006ApJ...645..792T}.

Equation \eqref{eq:profilenu} assumes that the ``laboratory frame''
(in which we are measuring the photon's frequency) is the same as the
fluid rest frame. In order to take account of the bulk fluid velocity, we 
replace the frequency in the lab frame, $x$, with the
frequency as seen by the fluid, 
\begin{equation} \bar{x} = x - \frac{\vv_\ro{b} \cdot
\nuv}{v_{\ro{th}}}.
\end{equation}

In practice, the photons are very likely to be emitted close to line
centre, and because the photons suffers a large number of resonant
scatterings, any ``memory'' of the initial frequency is quickly erased \citepalias{2006ApJ...649...14D}. Thus, we will
usually inject all photons at line centre in the fluid frame ($\bar{x}
= 0$).

\item \label{choosedist} The distance travelled depends on the optical
depth. The probability that the photon propagates a physical distance
corresponding to an optical depth between $\tau$ and $\tau + \dd \tau$
is $e^{-\tau} \dd \tau$. We choose an optical depth for the photon from this
distribution, 
\begin{equation} \tau = -\ln(R_3).
\end{equation}

To find the physical distance travelled, we perform the following
line integral, solving for $s_f$, 
\begin{equation} \label{eq:tauint} \tau_{x} = \int_0^{s_f}
\sigma_x(\rv(s)) ~ \dhi(\rv(s)) \df s, 
\end{equation} where $\rv(s) = \rv_i + \nuv s$ is the path travelled
by the photon, and $\sigma_x$ is the scattering cross-section of \lya
photons, 
\begin{equation} \sigma_x = f_{12} \pi c r_e \phi(x)
\end{equation} and $f_{12} = 0.4167$ is the \lya oscillator line
strength while $r_e = 2.81794 \times 10^{-13} ~\ro{cm}$ is the classical
electron radius. It is worth noting how the integrand depends on
position. The most obvious dependence is that of a spatially varying
density $\dhi(\rv)$. The dependence of $\sigma_\nu$ is two-fold. 
Firstly, in case of a spatially-varying bulk velocity, 
$\sigma_\nu(\phi(\bar{x}(\vv_\ro{b}(\rv))))$. Secondly, if $T$ depends
on position, then so will both $x$ and $a$ in $\phi(x)$ via $\DnuD$
and $v_{\ro{th}}$.

Once $s_f$ is found, the position of the next scattering is $\rv =
\rv_i + \nuv s_f$. If this is outside the H\textsc{i} region, then the
algorithm is terminated.

\item Next we choose the velocity of the scattering atom. Naively, one
might think that this step involves generating the three components of
the atom's velocity from the Maxwell-Boltzmann distribution. However, we are choosing
the velocity of an atom \emph{given} that it scatters a photon with
frequency $x$. We can therefore divide the velocity of the atom into one
component parallel ($v_{\shortparallel}$) and two components
perpendicular to the direction of motion of the photon ($v_{\bot 1}$,
$v_{\bot 2}$). The two perpendicular components do not alter the
frequency of the photon as seen by the atom, and are thus chosen from
a Maxwell-Boltzmann distribution. From Equations
\eqref{eq:profilenu}-\eqref{eq:voigt} it can be seen that the probability that a photon
with frequency $x$ scatters off an atom with velocity (along the
direction of propagation of the photon) between $v_{\shortparallel}$
and $v_{\shortparallel} + \dd v_{\shortparallel}$ is, 
\begin{equation} f(u_{\shortparallel}) \df u_{\shortparallel} =
\frac{a} {\pi}
\frac{e^{-u_{\shortparallel}^2}}{(x-u_{\shortparallel})^2 + a^2}
\frac{1}{H(a,x)} \df u_{\shortparallel}, 
\end{equation} where $u_{\shortparallel} =
v_{\shortparallel}/v_{\ro{th}}$. A scheme for generating random
numbers from this distribution is given in
\citet{2002ApJ...578...33Z}.

\item Now that we have calculated $\vv_a$, the velocity of the atom that scatters
the photon, we perform a Lorentz transform into the rest frame of the
atom. In this frame, we assume that the frequency of the scattered
photon differs from the frequency of the incident photon only by the
recoil effect \citep[][ pg. 196]{1979rpa..book.....R}. We choose a new
direction for the photon. Our code can incorporate either an isotropic
or a dipole distribution for the direction of the re-emitted photon. 
The results are insensitive to the choice. We then transform back into
the laboratory frame. At speeds much less than $c$ we have, for an
initial photon direction \nuv, and having chosen a final photon
direction $\nuv_f$, that the final frequency of the photon is,
\begin{equation} x_f = x - \frac{\nuv \cdot \vv_a}{v_{\ro{th}}} +
\frac{\nuv_f \cdot \vv_a}{v_{\ro{th}}} + \frac{h \DnuD}{2 k T}(\nuv
\cdot \nuv_f - 1).
\end{equation} In general, the final term, known as the recoil term
\citep{1971ApJ...168..575A}, is negligible for the modelling
in this paper.

We now return to step \ref{choosedist} and repeat it until the photon
escapes the H\textsc{i} region. 
\end{enumerate} 

Once the photon escapes the region, its properties (frequency, angle
of escape etc.) are recorded. We then return to step
\ref{generate} and generate another photon.

Our code incorporates the presence of a cosmic abundance of deuterium,
following the method presented in \citetalias{2006ApJ...649...14D}. 
Adding deuterium has only a minimal effect.

\subsection{Accelerating The Code}
Monte Carlo \lya RT codes like the one used here can be significantly accelerated
by skipping scatterings in the core of the line profile. We define the core-to-wing 
transition to occur at a critical frequency
\xcr, which is not the same as $x_c$. Whenever a photon is in the core
$|x| < \xcr$, we force it into the wing by choosing the scattering
atom's velocity to be large. We follow the method presented in
\citet{2002ApJ...567..922A}, \citetalias{2006ApJ...649...14D},
choosing the perpendicular components of the atom's velocity such that
$ u_{\bot 1}^2 + u_{\bot 1}^2 \ge \xcr^2$.

We choose the value of \xcr by requiring that, for a uniform sphere,
less than a fraction $f$ of the photons that emerge have $|x| <
\xcr$. Using the analytic solution for the emergent spectrum of a
sphere \citepalias[][Equation (9)]{2006ApJ...649...14D}, this gives, 
\begin{equation} \xcr = \left( \sqrt{\frac{54}{\pi^3}} a \tau_0
\tanh^{-1}f \right)^{\frac{1}{3}}, 
\end{equation} where $\tau_0$ is the line-centre optical depth from
the centre to the edge. We find that setting $f = 0.01$ up to
max$(\xcr) = 3 ~ ( \approx x_c)$ gives an acceptable compromise
between speed and accuracy. This procedure can accelerate the code by 
several orders of magnitude, e.g. for a static sphere with $\tau_0 = 10^7$, 
the code is accelerated by a factor of more than a thousand.

\subsection{Spherically Symmetric Shells} 
In cases where $\dhi$, $T$
and/or $\vv_\ro{b}$ have a complicated dependence on $\rv$, solving
for $s_f$ in Equation \eqref{eq:tauint} can be computationally
expensive. In spherical symmetry, we use a similar approach to
\citetalias{2006ApJ...649...14D}, dividing the sphere into 
shells of uniform density. Within each shell, the integral \eqref{eq:tauint} becomes
trivial, $\tau = s_f \sigma_x \dhi(r_\ro{shell})$. If the edge of a shell is
reached, the optical depth to the edge of the shell is subtracted from
$\tau$ and a new $s_f$ is calculated using the new value of
$r_\ro{shell}$. We space the shells so that each shell has equal
column density. We choose the number of shells to keep the frequency
dispersion within each shell small compared to the change in frequency
as the photon crosses a shell; 1000 shells are usually sufficient.


\section{The effect of temperature and velocity amplitude} \label{A:Tvampz}
In this section, we will consider the effect of altering the gas temperature 
and amplitude of the bulk velocity. For this section, the 
fiducial model is different to that in Section \ref{fidmod}:
$(z,M_v,c_0,f_e,v_\ro{amp},T) = (3,10^{11} \Msol, 3.5,1,
v_\ro{c},10^4 \ro{K})$.

\subsection{The effect of the velocity profile}
\begin{figure*} \centering
	\begin{minipage}[c]{0.74\textwidth} \centering 
 		\includegraphics[width=\textwidth]{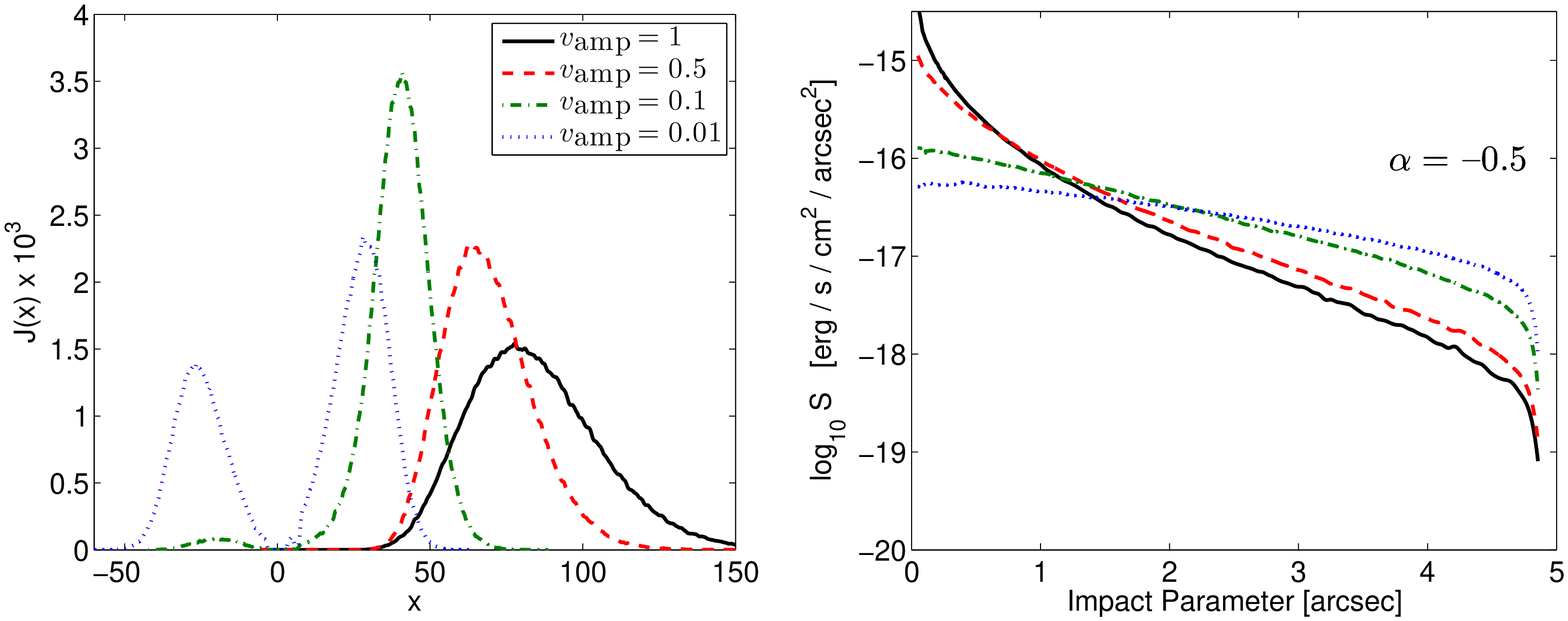}
		\includegraphics[width=\textwidth]{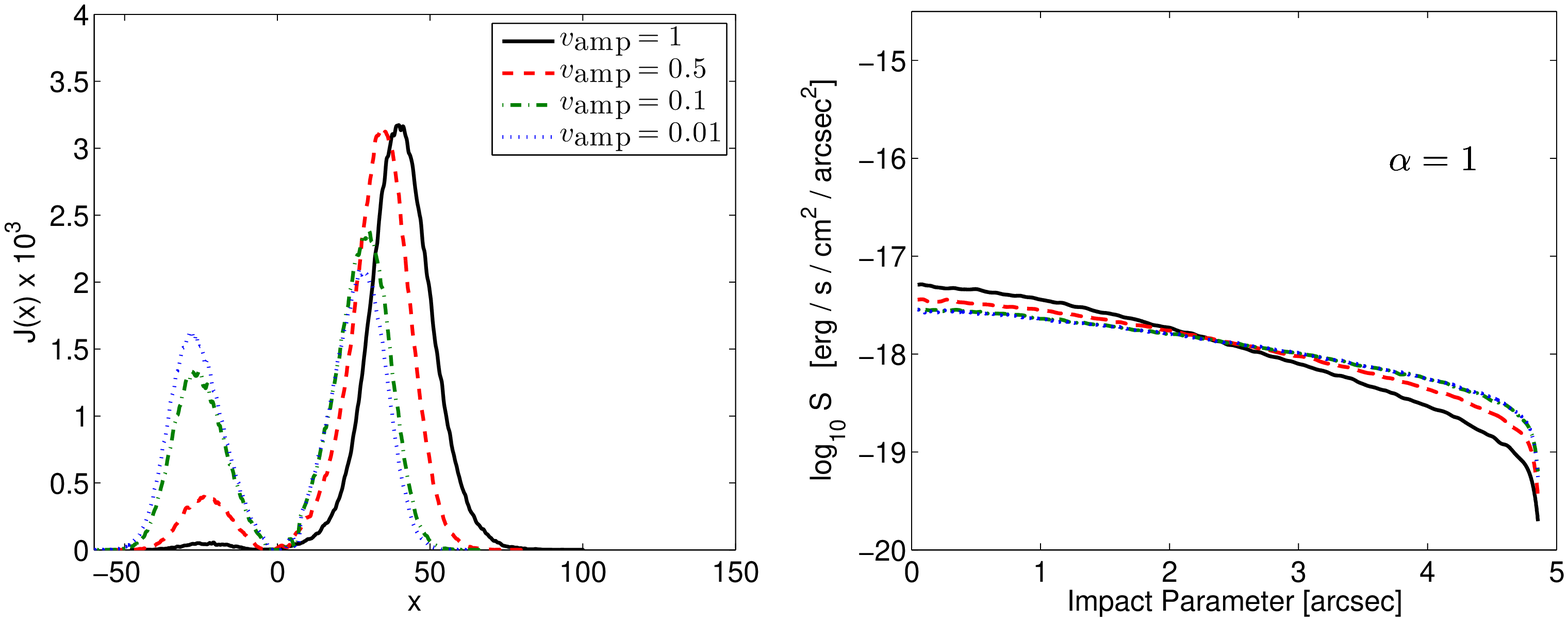}
	\end{minipage}
	\begin{minipage}[c]{0.25\textwidth} \centering 
 		\caption{Spectra (left panels) and surface brightness profiles (right panels) for
DLAs in halos with $v_\ro{amp}$ as given in the legend in units of
the virial velocity. The top panels are for $\alpha = -0.5$, for which 
$L_{\lya} = 1.1 \times 10^{44} \ergs$. The bottom panels are
for $\alpha = 1$, for which $L_{\lya} = 5.5 \times 10^{42} \ergs$. 
As the velocity decreases, the
spectrum begins to resemble the double-peaked static sphere
profile. The surface brightness profile flattens as the velocity
decreases --- where the bulk velocity is low, the photons must random
walk out of the Doppler core by scattering; they will not be given a
``free ride'' by the fluid flow. }
		\label{fig:Fbvaf}
	\end{minipage}
\end{figure*}

In this section, we consider the effect
of decreasing $v_\ro{amp}$. The results are shown in Figure
\ref{fig:Fbvaf}. Setting $v_\ro{amp}=0$ would result in $J(x)$ being
symmetric about $x=0$ (ignoring the small effects of deuterium and
recoil). This tendency is seen clearly in both of the left
plots. These plots also show that photons emerge bluer as the velocity
of infall increases. Photons gain energy from head-on collisions with
atoms, and the more energetic the atoms, the more energy is transferred
between photons and gas.

The surface brightness plots show that the emission becomes more
extended as the velocity is reduced. A rather flat surface brightness profile
is characteristic of a uniform, static \hi sphere. This is because the
bulk velocity can give photons a ``free ride'' through the halo,
Doppler shifting their frequency away from line centre in the fluid
frame without changing $x$ in the laboratory frame. In the $\alpha =
-0.5$ case, the peak surface brightness is reduced by almost 2 orders
of magnitude by this effect. It is also worth noting that the smaller
$v_\ro{amp}$ is, the less the dependence of the profile on $\alpha$.

\subsection{Temperature}
\begin{figure*} \centering
	\begin{minipage}[c]{0.74\textwidth} \centering 
		\includegraphics[width=\textwidth]{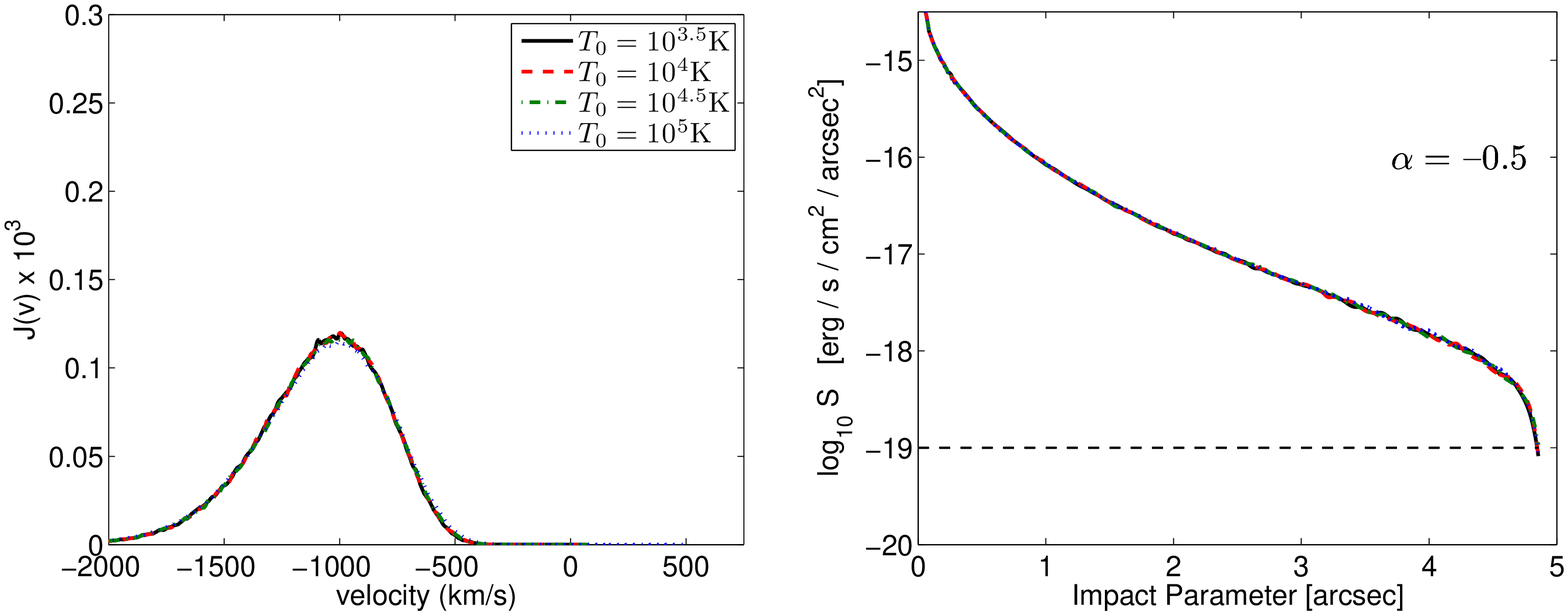}
		\includegraphics[width=\textwidth]{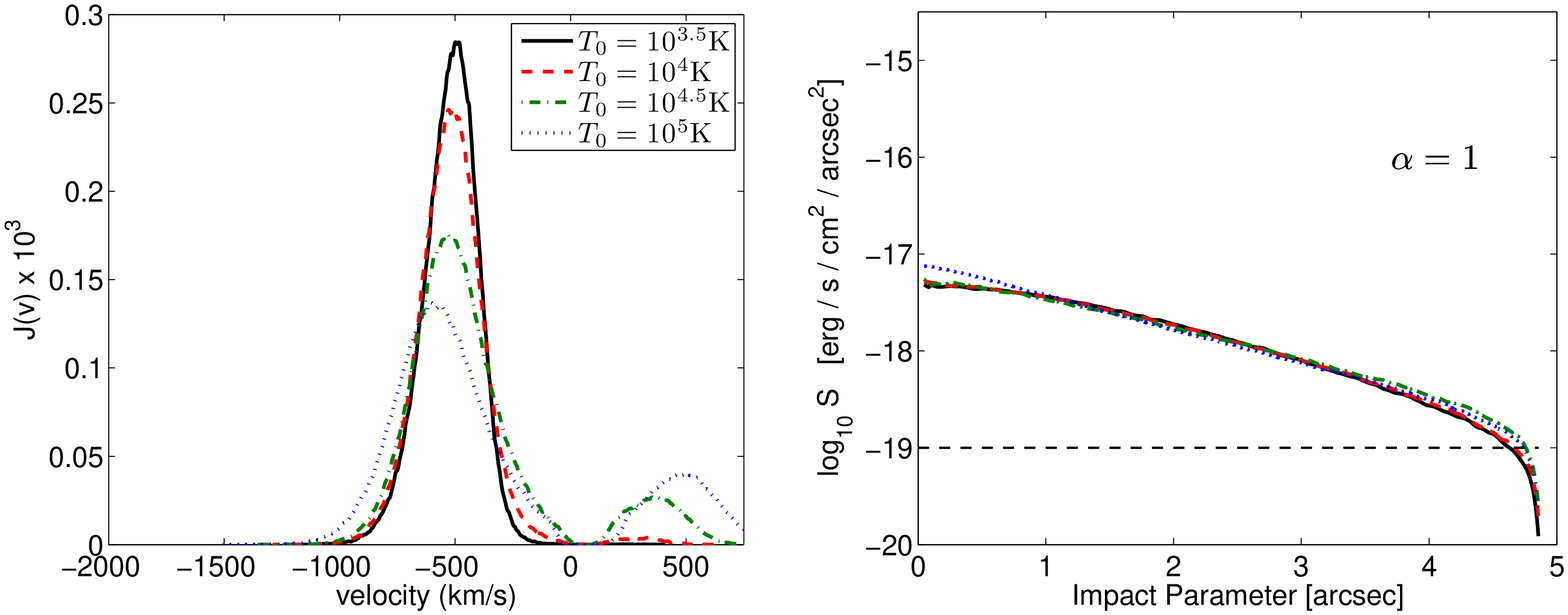}
	\end{minipage}
	\begin{minipage}[c]{0.25\textwidth} \centering 
 		\caption{Spectra and surface brightness profiles 
		as a function of \emph{velocity} for DLAs in haloes 
		with temperature $T_0$ as given in the legend. The top 
		panels are for $\alpha = -0.5$, for which 
		$L_{\lya} = 1.1 \times 10^{44} \ergs$. The bottom panels 
		are for $\alpha = 1$, for which $L_{\lya} = 5.5 \times 10^{42} \ergs$. 
		For the $\alpha = -0.5$ case, there is no dependence on the 
		temperature because the photon begins its flight with frequency 
		(in the fluid frame) in the scattering wings, and is unlikely to 
		return to the Doppler core. In the $\alpha = 1$ case, 
		the effect of temperature is minimal. The red peak disappears 
		at lower temperatures due to the corresponding increase in the 
		scattering cross-section, $\sigma_x \propto 1/\DnuD \propto 1/\sqrt{T}$, 
		when $x$ is small.}
		\label{fig:FbT}
	\end{minipage}
\end{figure*}

The effect of altering the temperature is shown in Figure \ref{fig:FbT}. 
Note that the spectrum is given as a function of velocity 
($v \equiv -x ~ v_{\ro{th}}$), as $x$ itself is temperature dependent 
($\propto 1/\sqrt{T}$). For the $\alpha = -0.5$ case, there is no 
dependence on the temperature because the photon begins its flight 
with a frequency (in the fluid frame) in the scattering wings, and 
is unlikely to return to the Doppler core. Remember that the 
scattering cross-section does not depend on $T$ in the Lorentz 
wing of the profile. In the $\alpha = 1$ case, the effect of 
temperature is minimal. The red peak disappears at lower temperatures 
due to the corresponding increase in the scattering cross-section, 
$\sigma_x \propto 1/\DnuD \propto 1/\sqrt{T}$, when $x$ is small. 


\bsp 
\label{lastpage}

\end{document}